\documentclass[aps,pre,nofootinbib,showkeys,notitlepage]{revtex4-1}
\pdfoutput=1
\usepackage[colorlinks,bookmarksopen,bookmarksnumbered,citecolor=red,urlcolor=red]{hyperref}

\usepackage{moreverb}

\usepackage[utf8]{inputenc}
\usepackage[english]{babel}
\usepackage{amsfonts,amssymb,amsbsy,amsmath}
\usepackage{graphicx}
\usepackage{placeins}
\usepackage{caption}
\usepackage{subcaption}
\usepackage{natbib}

\newcommand{\trans}{\mathsf{T}}
\newcommand{\tens}[1]{\boldsymbol{\rm #1}}
\newcommand{\D}{\mathrm{d}}
\newcommand{\I}{{i\mkern1mu}}

\begin{document}

\title{Stochastic parameterization of subgrid-scale processes: A review of recent physically-based approaches}

\author{Jonathan Demaeyer}
\email{E-mail: Jonathan.Demaeyer@meteo.be}
\author{St\'{e}phane Vannitsem}
\affiliation{Royal Meteorological Institute of Belgium, Avenue Circulaire, 3, 1180 Brussels, Belgium}

\begin{abstract}
  We review some recent methods of subgrid-scale parameterization used in the context of climate modeling. These methods are developed to take into account (subgrid) processes playing an important role in the correct representation of the atmospheric and climate variability. We illustrate these methods on a simple stochastic triad system relevant for the atmospheric and climate dynamics, and we show in particular that the stability properties of the underlying dynamics of the subgrid processes has a considerable impact on their performances.
\end{abstract}

\maketitle

\section{Introduction}
\label{sec:intro}

From a global point of view, the Earth system is composed of a myriad of different interacting components. These components can be regrouped in compartments like the atmosphere, the hydrosphere, the lithosphere, the biosphere and the cryosphere~\cite{O2001}\footnote{More recently, a new compartment have appeared, whose effect is not negligible at all and which is not predictable nor descriptive by evolution equations, namely the impact of the human activities.}. Those compartments play a role on different timescales from seconds to ice ages. In this perspective, the resulting Earth's climate is a ``concert'' at which each compartment seems to play its own partition with its own tempo. Their respective contribution to the total variability of an observable, say e.g. the global temperature, is, however, the outcome of complex interactions between the different components, leading to an emergent dynamics far from the one that could be generated by a linear additive superposition principle~\cite{NN1981,NN2012}.

A paradigmatic example is provided in the work of Hasselmann, detailed in his seminal paper of 1976, which states precisely that the slowly evolving components of the climate system, besides their own dynamics due their own physical processes, also integrate the impact of the faster components~\cite{H1976}. Hasselmann describes this process using the analogy of the Brownian motion where a macro-particle in a liquid integrates the effect of the collisions with the fluid's micro-particles, leading to the erratic trajectory of the former. The interest of this framework is that it provides a natural description of the ``red noise'' spectral density observed in most climatic records and observations~\cite{Ga2002,LS2013}. Subsequently, during the following decade, stochastic modeling for meteorology and climatology became an important research topic~\cite{L1977,FH1977,F1979,FM1979,LTH1980,NN1981,N1981,N1982,P1989} before falling into disuse in what has been described as a ``lull'' of work in this field~\cite{AIW2003}. However, during that period, the idea that correct parameterizations of sub-grid processes are important to improve climate and weather models gained popularity~\cite{PM1994,P1996,NSP1997}. Stochastic parameterizations for the ``turbulent'' closure in 2-D large-eddy simulations on the sphere have also been considered~\cite{FD1997,F1999}. It led recently to the implementation of stochastic schemes to correct the model errors~\cite{N2003,N2004} made in large numerical weather prediction (NWP) models~\cite{BMP1999,S2005}, improving the reliability of probabilistic forecasts and correcting partially their variability~\cite{N2005,Da2009}. The relation between multiplicative noise and the non-Gaussian character observed in some geophysical variables has also been considered~\cite{SNPS2005,SP2015}, as well as stochastic models for the climate extremes~\cite{S2013}.

Since the beginning of the 21st century, a revival of the interest in stochastic parameterization methods have occurred, due to the availability of new mathematical methods to perform the stochastic reduction of ODEs systems. Almost simultaneously, a rigorous mathematical framework for the Hasselmann ``program'' was devised~\cite{K2001,A2001,K2003} and a new method based on the singular perturbation theory of Markov processes~\cite{MTV2001} was proposed. The latter approach is currently known as the Majda-Timofeyev-Vanden-Eijnden (MTV) method. Both methods have been tested and implemented successfully in geophysical models~\cite{FMV2005,CKM2011,AIW2003,V2014}. The revival of the Hasselmann program has also stressed the need to consider the occurrence of very rare events triggered by the noise that allow for the solutions of the system to jump from one local attractor to another one~\cite{A2001}. Such events display recurrence timescales that are few orders greater than the timescale of the climate variables considered, and thus induce transitions between different climates. The statistics of these transitions is then given by the so-called \emph{Large Deviations} theory~\cite{FW1984} (for recent developments on this matter, see Ref.~\cite{BGTV2016}). In addition to these two methods, the modeling of the effects of subgrid scale through conditional Markov chain has been considered~\cite{CV2008} and recently, new stochastic parameterization techniques have been proposed, based on an expansion of the backward Kolmogorov equation~\cite{A2015} and on the Ruelle response theory~\cite{WL2012}. The latter has been tested on a simple coupled ocean-atmosphere model~\cite{DV2016}, on stochastic triads~\cite{WDLA2016} and on an adapted version of the Lorenz'96 model~\cite{VL2016}.

This renewal of interest for stochastic modeling and reduction methods illustrates how fruitful was the original idea of Hasselmann. However, in view of the availability of several possible approaches, one might wonder about their efficiency in different situations. Indeed, depending on the specific purpose that it needs to fulfill, some parameterizations might perform better than others. The present review aims to shed some light on these questions and to illustrate some of the aforementioned parameterization methods on a simple model for which most of the calculations can be made analytically.

In Section~\ref{sec:param_prob}, we will present the general framework in which the problem of stochastic parameterizations is posed. In Section~\ref{sec:mod_param}, we present the different parameterizations that we shall consider for the analysis model. The stochastic triad model used here and the comparison are presented in Section~\ref{sec:appli}. Finally, the conclusions are given in Section.~\ref{sec:conclu}.

\section{The parameterization problem}
\label{sec:param_prob}

Consider the following ordinary differential system of equations (ODEs):
\begin{equation}
  \label{eq:gen_sys}
  \dot Z = T(Z)
\end{equation}
where $Z\in \mathbb{R}^d$ is a set of variables relevant for the problem under interest for which the tendencies $T(Z)$ are known. And suppose that one wants to separate this set of variables into two different subset $Z=(X,Y)$, with $X\in \mathbb{R}^m$ and $Y\in \mathbb{R}^n$. In general, such a decomposition is made such that the subset $X$ and $Y$ have strongly differing \emph{response times} $\tau_Y \ll \tau_X$~\cite{AIW2003}, but we will assume here that this constraint is not necessarily met. System~(\ref{eq:gen_sys}) can then be expressed as:
\begin{equation}
  \label{eq:dec_gen_sys}
  \left\{
  \begin{array}{lcl}
    \dot{X} & = & F(X,Y) \\%= F_X(X)+  \Psi_X(X,Y) \\
    \dot{Y} & = & H(X,Y) %= F_Y(Y)+  \Psi_Y(X,Y) \\
  \end{array}
  \right.
\end{equation}
The timescale of the $X$ sub-system is typically (but not always) longer than the one of the $Y$ sub-system, and it is often materialized by a parameter $\delta=\tau_Y/\tau_X \ll 1$ in front of the time derivative $\dot Y$. The $X$ and the $Y$ variables represent respectively the resolved and the unresolved sub-systems. The resolved sub-system is the part of the full system that we would like to simulate, i.e. generate explicitly and numerically its time-evolution. The general problem of model reduction consists thus to approximate the resolved component $X$ as accurately as possible by obtaining a closed equation for the system $X$ alone~\cite{AIW2003}. The term ``accurately'' here can have several meanings, depending on the kind of problem to solve. For instance we can ask that the closed system for $X$ has statistics that are very close to the ones of the $X$ component of system~(\ref{eq:dec_gen_sys}). We can also ask that the closed system trajectories remains as close as possible to the trajectories of the full system for long times.

In general a parameterization of the sub-system $Y$ is a relation $\Xi$ between the two sub-systems:
\begin{equation}
  \label{eq:param}
  Y=\Xi(X,t)
\end{equation}
which allows to effectively close the equations for the sub-system $X$ while retaining the effect of the coupling to the $Y$ sub-system.

The problem of the model reduction is not new, and was considered first in celestial mechanics. Famous mathematician have considered it and contributed to what is known nowadays as the theory of averaging~\cite{SV1985} and which forms the first step of the Hasselmann program~\cite{A2001}. The mathematical framework was set in the 1960s by the influential contribution of Bogoliubov and Mitropolski~\cite{BM1961}. However, this averaging technique is a deterministic method which does not take into account the deviations from the average. The proposition of Hasselmann was thus to take into account these deviations by considering stochastic parameterization where the relation~(\ref{eq:param}) can be considered in a statistical sense. In that framework, the $Y$ sub-system and its effect on the sub-system $X$ can be considered as a stochastic process, which possibly depends upon the state of the $X$ sub-system. Different methods to achieve this program are now discussed in Section~\ref{sec:mod_param}.
\section{The parameterization methods}
\label{sec:mod_param}
Let us now write system~(\ref{eq:dec_gen_sys}) as:
\begin{equation}
  \label{eq:dec_gen_sys_spec}
  \left\{
  \begin{array}{lcl}
    \dot{X} & = & F_X(X)+  \Psi_X(X,Y) \\
    \dot{Y} & = & F_Y(Y)+  \Psi_Y(X,Y) 
  \end{array}
  \right.
\end{equation}
where the coupling and the intrinsic dynamics are explicitly specified.
In the present work, following the Hasselmann program, we shall focus on parameterizations that are defined in terms of stochastic processes. We will consider methods based
\begin{itemize}
\item on the Ruelle response theory~\cite{WL2012,DV2016,WDLA2016}.
\item on the singular perturbation theory of Markov processes~\cite{MTV2001,FMV2005}.
\item on the Hasselmann averaging methods~\cite{K2003,AIW2003,CKM2011,V2014}.
\item on empirical methods~\cite{AMP2013}.
\end{itemize}
All these parameterizations can be written in the following form:
\begin{equation}
  \label{eq:Xparam}
  \dot X = F_X(X) + G(X,t) + \tens{\sigma}(X) \cdot \tilde\xi(t)
\end{equation}
where the matrix $\tens\sigma$, the deterministic function $G$ and the random processes $\tilde\xi(t)$ have to be determined. The mathematical definition of these quantities obtained through averaging procedure and the measure being used to perform the averaging are usually both differing between the methods. These different choices are rooted in their different underlying hypothesis, as it will be discussed below. Specifically, the response theory method uses the measure of the uncoupled unresolved sub-system $\dot Y = F_Y(Y)$, the singular perturbation method uses the measure of the perturbation, and the averaging methods use the measure of the full unresolved sub-system $\dot Y = H(X,Y)$ with $X$ considered as ``frozen'' (constant). Finally, the empirical methods use in general the output of the full unresolved $Y$ sub-system, conditional or unconditional on the state $X$.

% Notwithstanding the conceptual differences between the methods, the choice of the measure used to compute the averages depends in general depending on the problem considered. This choice can be motivated by the technical difficulties or by the methodological assumptions considered. For instance, in Ref.~\cite{FMV2005}, the singular perturbation method is used with the measure of the full dynamics, assuming that it approximates well the measure of the perturbation. In this respect, the example that we shall consider in the following section was selected because the different measures involved can be computed analytically.
In the rest of the section, we shall describe more precisely each of the above methods.
\subsection{The method based on response theory} 
\label{sec:WL}
This method is based on the Ruelle response theory~\cite{R1997,R2009} and was proposed by Wouters and Lucarini~\cite{WL2012,WL2013}.
In this context, system~(\ref{eq:dec_gen_sys_spec}) must be considered as two intrinsic sub-dynamics for $X$ and $Y$ that are weakly coupled. The response theory quantifies the contribution of the ``perturbation'' $\Psi_X$, $\Psi_Y$ to the invariant measure\footnote{The theory assumes that for the system under consideration, a SRB measure~\cite{Y2002} exists (e.g. an Axiom-A system).} $\tilde\rho$ of the fully coupled system~(\ref{eq:dec_gen_sys_spec}) as:
\begin{equation}
  \label{eq:pertrho}
  \tilde\rho = \rho_0 + \delta_\Psi \rho^{(1)} + \delta_{\Psi,\Psi} \rho^{(2)} + O(\Psi^3)
\end{equation}
where $\rho_0$ is the invariant measure of the uncoupled system which is also supposed to be an existing, well defined SRB measure.
As shown in~\cite{WL2012}, this theory gives the framework to parameterize the effect of the coupling on the component $X$. The parameterization is based on three different terms having a response similar, up to order two, to the couplings $\Psi_X$ and $\Psi_Y$:
\begin{equation}
  \label{eq:WLres}
  \dot X = F_X(X) + M_1(X) + M_2(X,t) + M_3(X,t)
\end{equation}
where
\begin{equation}
  \label{eq:M1def}
  M_1(X) = \Big\langle \Psi_X(X,Y) \Big\rangle_{\rho_{0,Y}}
\end{equation}
is an averaging term. $\rho_{0,Y}$ is the measure of the uncoupled system $\dot Y=F_Y(Y)$. The term $M_2(X,t)=\sigma_R(X,t)$ is a correlation term:
\begin{equation}
  \label{eq:gdef}
  \Big\langle \sigma_R(X,t) \otimes \sigma_R(X,t')\Big\rangle = \tens g (X,s) = \Big \langle \Psi_X^\prime(X,Y)  \otimes \Psi_X^\prime\big(\phi^s_X(X),\phi^s_Y(Y))\big) \Big \rangle_{\rho_{0,Y}}
\end{equation}
where $\otimes$ is the outer product, $\Psi_X^\prime(X,Y)=\Psi_X(X,Y)-M_1(X)$ is the centered perturbation and $\phi^s_X$, $\phi^s_Y$ are the flow of the uncoupled system $\dot X=F_X(X)$ and $\dot Y = F_Y(Y)$. The process $\sigma_R$ is thus obtained by taking the square root of the matrix $\tens g$, which is here accomplished by a decomposition of $\Psi_X^\prime$ on a proper basis~\cite{WL2012}. The $M_3$ term is a memory term:
\begin{equation}
  M_3(X,t) = \int_0^\infty \D s \, h(X(t-s),s). \label{eq:M3def}
\end{equation}
involving the memory kernel
\begin{equation}
  \label{eq:M3gen}
  h (X,s) = \Big\langle \Psi_Y(X,Y) \cdot \tens\nabla_Y \Psi_X\big(\phi^s_X(X),\phi^s_Y(Y)\big) \Big\rangle_{\rho_{0,Y}}
\end{equation}

All the averages are thus taken with $\rho_{0,Y}$, the invariant measure of the unperturbed system $\dot{Y} = F_Y(Y)$. This particular choice of the measure is due to the perturbative nature of the method and simplify the averaging procedure in many case. The terms $M_1$, $M_2$ and $M_3$ are derived~\cite{WL2012} such that their responses up to order two match the response of the perturbation $\Psi_X$ and $\Psi_Y$. Consequently, this ensures that for a \emph{weak coupling}, the response of the parameterization~(\ref{eq:WLres}) on the observables will be approximately the same as the coupling.

The advantages of this simplified averaging procedure (by using $\rho_{0,Y}$) should be tempered by the additional cost induced by the computation of the memory term, the latter implying that this parameterization is a non-Markovian one~\cite{CLW2015}. However, the integral~(\ref{eq:M3def}) in this memory term must only be evaluated from $0$ up to the timescale $\tau_Y$ of the fast variable, due to the exponential decrease of the integrand. Moreover, in some cases, this non-Markovian parameterization can be effectively replaced by a Markovian one~\cite{WDLA2016}.

\subsection{Singular perturbation theory method}
\label{sec:MTV}
Singular perturbation methods  were developed in the 1970s for the analysis of the linear Boltzmann equation in some asymptotic limit~\cite{G1969,EP1975,P1976,MTV2001}. Here, these methods are applicable if the problem can be cast into a Fokker-Planck equation. The procedure described in Ref.~\cite{MTV2001} requires assumptions on the timescales of the different terms of system~(\ref{eq:dec_gen_sys_spec}). In term of the small parameter $\delta=\tau_Y/\tau_X$ defined in Section~\ref{sec:param_prob}, the fast variability of the unresolved component $Y$ are considered of order $O(\delta^{-2})$ and modeled as an Ornstein-Uhlenbeck process. The Markovian nature of the process defined by Eq.~(\ref{eq:dec_gen_sys_spec}) and its singular behavior in the limit of an infinite timescale separation ($\delta\to 0$) allow then to apply the method. % In this framework, we will consider that the system~(\ref{eq:dec_gen_sys_spec}) can be written as:
% \begin{equation}
%   \label{eq:MTV_sys}
%     \left\{
%   \begin{array}{lcl}
%     \dot{X} & = & F(X,Y) = F_X(X)+  \frac{1}{\delta}\Psi_X(X,Y) \\
%     \dot{Y} & = & H(X,Y) = \frac{1}{\delta^2} F_Y(Y)+  \frac{1}{\delta}\Psi_Y(X,Y) \\
%     \end{array}
%   \right
% \end{equation}

More specifically, the parameter $\delta$ serves to distinguish terms with different timescales and is then used as a small perturbation parameter. In this setting, the backward Fokker-Planck equation reads~\cite{MTV2001}:
\begin{equation}
  \label{eq:FPback}
  -\frac{\partial \rho^\delta}{\partial s} = \left[ \frac{1}{\delta^2} \mathcal{L}_1 +\frac{1}{\delta}\mathcal{L}_2 + \mathcal{L}_3 \right] \rho^\delta
\end{equation}
where the function $\rho^\delta(s,X,Y|t)$ is defined with the final value problem $f(X)$: $\rho^\delta(t,X,Y|t) =f(X)$. The function $\rho^\delta$ can be expanded in term of $\delta$ and inserted in Eq.~(\ref{eq:FPback}). The zeroth order of this equation $\rho^0$ can be shown to be independent of $Y$ and its evolution given by a closed, averaged backward Fokker-Planck equation~\cite{K1973}:
\begin{equation}
  \label{eq:avFPback}
  -\frac{\partial \rho^0}{\partial s} = \bar{\mathcal{L}} \rho^0
\end{equation}
This equation is obtained in the limit $\delta\to 0$ and gives the sought limiting, averaged process $X(t)$.
Note that this procedure does not necessarily require the presence of the explicit small parameter $\delta$ in the original equation~(\ref{eq:dec_gen_sys_spec}). Since $\delta$ disappears from Eq.~(\ref{eq:avFPback}), one can simply use the parameter to identify the fast terms to be considered, and eventually consider $\delta=1$~\cite{FMV2005}.

The parameterization obtained by this procedure is given by~\cite{FMV2005}:
\begin{equation}
  \label{eq:majda_params}
  \dot X = F_X(X) +G(X) + \sqrt{2}\, \tens\sigma_{\rm MTV}(X) \cdot \tilde\xi(t)
\end{equation}
with
\begin{eqnarray}
  \label{eq:majda_def}
  G(X) &= &\int_0^\infty \, \D s \, \Big[\big\langle \Psi_Y(X,Y)\cdot \nabla_Y \Psi_X(X,\phi^s_Y(Y))\big\rangle_{\tilde\rho} \nonumber \\
    & & \qquad \qquad + \big\langle \Psi_X(X,Y)\cdot \nabla_X \Psi_X(X,\phi^s_Y(Y))\big\rangle_{\tilde\rho} \Big] \\
  \tens \sigma_{\rm MTV}(X) & = & \left(\int_0^\infty \, \D s \, \big\langle \Psi_X^\prime(X,Y) \, \Psi_X^\prime(X,\phi^s_Y(Y)) \big\rangle_{\tilde\rho}\right)^{1/2} \label{eq:SMTVdef}
\end{eqnarray}
with the same notation as in the previous subsection.
The measure $\tilde\rho$ is the measure of the $O(\delta^{-2})$ perturbation, i.e. the source of the fast variability of the unresolved $Y$ component. This measure thus depends on which terms of the unresolved component are considered as ``fast'', and some assumptions should here be made. For instance, it is customary to consider as the fast terms the quadratic terms in $Y$ and replaced them by Ornstein-Uhlenbeck processes whose measures is used to compute the averages~\cite{MTV2001,FMV2005}.

Finally, if one assume that the source of the fast variability in the sub-system is given by the ``intrinsic'' term $F_Y(Y)$ (such that $\tilde\rho=\rho_{0,Y}$) and if the perturbation $\Psi_X$ only depends on $Y$, this parameterization is simply given by the integration of the function $g(s)$ and $h(X,s)$ of the response theory parameterization given by Eqs.~(\ref{eq:gdef}) and~(\ref{eq:M3gen}). This can be interpreted as an averaging of the latter parameterization when the timescale separation is infinite and $X$ can thus be considered as constant over the timescale of the integrand. Therefore, $M_2$ can be modeled as a white noise and the memory term is Markovian.

\subsection{Hasselmann averaging method}
\label{sec:averaging}
Since the initial work of Hasselmann in the 1970s~\cite{H1976}, various approaches have been considered to average directly the effects of the ``fast'' evolving variables on the ``slow'' ones. These methods assume in general a sufficient timescale separation between the resolved and unresolved components of the systems, and a direct average can be performed as,
\begin{equation}
  \label{eq:dave}
  \dot X = \bar F(X) = \big\langle F(X,Y) \big\rangle_{\rho_{Y|X}}
\end{equation}
where $\rho_{Y|X}$ is the measure of the system 
\begin{equation}
  \label{eq:Yevol}
  \dot Y = H(X,Y)
\end{equation}
 conditional on the value of $X$. In this approach, $X$ is thus viewed as a constant parameter for the unresolved dynamics. In other words, this particular framework assumes that since $X$ is slowly evolving with respect to the typical timescale of $Y$, it can be considered as ``frozen'' while $Y$ evolves. With some rigorous assumptions, this approach has been mathematically justified~\cite{K2003} and applied successfully to idealized geophysical models~\cite{AIW2003} with non-trivial invariant measures. In the same vein, an approximation has been proposed in Ref.~\cite{A2013} for the average~(\ref{eq:dave}), assuming that $F$ is at most quadratic,
 \begin{equation}
   \label{eq:daveapp}
   \big\langle F(X,Y) \big\rangle_{\rho_{Y|X}} =  F(X,\bar Y(X)) + \frac{1}{2} \frac{\partial^2 F}{\partial Y^2}(X,\bar Y(X)) : \tens\Sigma(X)
 \end{equation}
where ``:'' means the element-wise matrix product with summation and where
\begin{eqnarray}
  \label{eq:bardef}
   \bar Y(X) & = & \langle Y \rangle_{\rho_{Y|X}} \\
   \tens\Sigma(X) & = & \langle (Y-\bar Y(X) \otimes (Y-\bar Y(X)) \rangle_{\rho_{Y|X}} \quad .
\end{eqnarray}
The approximation to the second order is particularly well suited for the application to atmospheric and climate flows for which the quadratic terms are usually the main non-linearities associated with the advection in the system.

In Ref.~\cite{A2013}, an approach based on the fluctuation-dissipation theorem was proposed to estimate the mean state $\bar Y(X)$ and the covariance matrix $\tens\Sigma(X)$. % In the present case, this approach is not needed since $\bar Y(X)=0$ for every state $X$, and $\tens\Sigma(X)$ solves the equation:
% \begin{equation}
%   \label{eq:stat_covx}
%   \tens T(X)\cdot \tens\Sigma(X) + \tens\Sigma(X) \cdot \tens T(X)^\trans = -\frac{1}{\delta^2} \tens B_Y  \cdot \tens B_Y^\trans
% \end{equation}
% with
% \begin{equation}
%   \label{eq:T_def}
%   \tens T(X) = \frac{1}{\delta^2}\tens{A}+\frac{\varepsilon}{\delta} X\, \tens V \quad .
% \end{equation}

The deterministic parameterization~(\ref{eq:dave}) can be recast in a stochastic parameterization following the same principle. Such a parameterization has been derived in Refs.~\cite{AIW2003,A2015} and reads
\begin{equation}
  \label{eq:abramov_stoch_def}
  \dot X =  \bar F(X) + \sigma_{\rm A} (X) \cdot \xi(t)
\end{equation}
with 
\begin{equation}
  \label{eq:ave_sigma}
  \sigma_{\rm A} (X) = \left(2 \int_0^\infty \, \D s \, \left\langle \big(F(X,\phi_{Y|X}^s Y)-\bar F(X)\big)\,\big(F(X,Y)-\bar F(X)\big)\right\rangle_{\rho_{Y|X}}\right)^{1/2}
\end{equation}
where $\phi_{Y|X}^s$ is the flow of the system~(\ref{eq:Yevol}) for $X$ constant (``frozen''). A drawback of such an approach is that it requires that the measure $\rho_{Y|X}$ exists and is well-defined (ideally a SRB measure~\cite{AIW2003}). Such a requirement may not be always fulfilled, like for instance if the fast system conditional on the state $X$ is unstable and does not possess any attractor (see Section~\ref{sec:appli} for an example).

\subsection{Empirical methods}
\label{sec:empiric}
The empirical methods are generally based on the statistical analysis of the timeseries $Y$ of the full system~(\ref{eq:dec_gen_sys_spec}). Many procedures exist as discussed in Section~\ref{sec:intro} but we will consider here a method based on state dependent AR(1) processes proposed in Ref.~\cite{AMP2013}. In this case, a timeserie $r(t)$ of the coupling part $\Psi_X$ of the $X$ tendency must first be computed with~(\ref{eq:dec_gen_sys_spec}). The parameterization is then given by
\begin{equation}
  \label{eq:emp_par}
  \dot X = F_X(X) + \mathcal{U}(X)
\end{equation}
with 
\begin{equation}
  \label{eq:emp_add}
  \mathcal{U}(X) = \mathcal{U}_{\det}(X) + e(X(t),t) \quad .
\end{equation}
The function $\mathcal{U}_{\det}(X)$ represents the deterministic part of the parameterization and is obtained by a least-squares fit of the timeserie $r(t)$ versus the timeserie $X(t)$ with the cubic function $\mathcal{U}_{\det}(X) = p_0 + p_1 \, X + p_2 \, X^2 + p_3 \, X^3$.
The ``stochastic'' part $e(X(t),t)$ is then given by the following state-dependent AR(1) process
\begin{equation}
  \label{eq:emp_ar1}
  e(X(t),t)= \phi \, \frac{\sigma_e(X(t))}{\sigma_e(X(t-\Delta t))} \, e(X(t-\Delta t),t-\Delta t) + \sigma_e(X(t)) \, (1-\phi^2)^{1/2} \, z(t)
\end{equation}
where $z(t)$ is a standard Gaussian white noise process. The parameters of the process $e$ are determined by considering the residual timeserie $r(t) - \mathcal{U}_{\det}(X(t))$ to compute the lag-1 autocorrelation $\phi$ and the state dependent standard deviation $\sigma_e(X)$ which is modeled as $\sigma_e(X) = \sigma_0 + \sigma_1 \, | X |$ with the parameters $\sigma_0$ and $\sigma_1$ given by a binning procedure. The parameter $\Delta t$ is the time step of integration of Eq.~(\ref{eq:emp_par}). Other empirical parameterizations have been proposed by the authors of Ref.~\cite{AMP2013}, notably one with the function $\mathcal{U}(X) = (1+e(t))\, \mathcal{U}_{\det}(X)$ which resembles the SPPT\footnote{Acronym for \emph{Stochastically Perturbed Parameterization Tendencies Scheme}.} parameterization used in the ECMWF\footnote{Acronym for \emph{European Center for Medium-Range Weather Forecasts}.} Numerical Weather Prediction model~\cite{BMP1999}. However, the study shows no substantial differences with the parameterization~(\ref{eq:emp_add}).

\section{Applications and results}
\label{sec:appli}

In this section, we will illustrate the various parameterizations described in Section~\ref{sec:mod_param} to the following example
\begin{equation}
  \label{eq:triad_sys}
  \left\{
  \begin{array}{lcl}
      \dot{X} & = & - D\, X + q\, \xi(t) + \frac{\varepsilon}{\delta} Y^\trans \cdot \tens{C} \cdot Y\\
      \dot{Y} & = & \frac{1}{\delta^2} \Big(\tens{A}\cdot Y + \delta \, \tens B_Y \cdot \xi_Y(t)\Big) + \frac{\varepsilon}{\delta} X \, \tens V \cdot Y
    \end{array}
  \right.
\end{equation}
where $D > 0$, $q >0$ and 
\begin{equation}
  Y=\left[
    \begin{array}{c}
      y_1 \\ y_2
    \end{array}
    \right] \quad .
\end{equation}
with $X$, $y_1$, $y_2 \in\mathbb{R}$.
The matrices involved are defined as
\begin{equation}
  \label{eq:mat_def}
  \tens C = \left[
    \begin{array}{cc}
      0 & B \\ B & 0
    \end{array}
  \right] \quad , \quad
  \tens A = \left[
    \begin{array}{cc}
      -a & \beta \\ -\beta & -a
    \end{array}
  \right] \quad , \quad
  \tens V = \left[
    \begin{array}{cc}
       0 & B_1 \\ B_2 & 0
    \end{array}
  \right] \quad \mathrm{and} \quad
  \tens B_Y = \left[
    \begin{array}{cc}
      q_Y & 0 \\ 0 & q_Y
    \end{array}
  \right]
\end{equation}
with $a,\beta,q_Y > 0$. The process $\xi(t)$ and $\xi_Y(t)$ are uncorrelated standard Gaussian white noise processes. 

The $X$ and $Y$ variables represent respectively the resolved and the unresolved sub-systems. The parameter $\delta > 0$ quantify the time-scale separation of the terms of the tendencies of the two components, with the three timescales $O(1)$,$O(\delta^{-1})$ and $O(\delta^{-2})$ as in Ref.~\cite{MTV2001} (see also Subsection~\ref{sec:MTV}). Additionally, the parameter $\varepsilon >0$ controls the coupling strength between the two sub-systems. In this setup the coupling are thus proportional to the ratio $\varepsilon/\delta$, and therefore the characterization of the coupling as ``weak'' depends directly on the time-scale separation.

The deterministic part of Eq.~(\ref{eq:triad_sys}) is a well known model called a triad and encountered in fluid dynamics~\cite{W1992,OK1992,SW1999}, and in simplified geophysical flows e.g.~\cite{MTV2001,WDLA2016}. Due to the presence of invariant manifolds, its mathematical structure can be found in higher-order model. See Ref.~\cite{DV2016} for an example of such structure in the framework of a coupled ocean-atmosphere model. In the present study, the interest of the stochastic triad model~(\ref{eq:triad_sys}) is that, $H(X,Y)$ being linear in $Y$, the measure $\rho_{0,Y}$ and $\rho_{Y|X}$ can be analytically computed since both $\dot Y = F_Y(Y)$ and $\dot Y = H(z,Y)|_{z=X}$ are two-dimensional Ornstein-Uhlenbeck processes. Therefore, for this simple case, the set of methods proposed in the previous section can be applied exactly without resorting to a binning procedure of the output of the $Y$ sub-system\footnote{Except for the empirical methods which by definition use this kind of procedures.}.

As energy conservation is a rule in physical systems in the absence of dissipation and fluctuations, we will adopt this rule for the current system. With the dissipation and stochastic terms set to zero, the system~(\ref{eq:triad_sys}) conserves the ``energy'' $(X^2+y_1^2+y_2^2)/2$ if the coefficient $B$, $B_1$ and $B_2$ are chosen such that~\cite{SW1999,MTV2001}
\begin{equation}
  \label{eq:consE}
  2 B+B_1+B_2=0 \quad .
\end{equation}
It allows for the following configurations of their signs : $(+,-,-)$, $(+,+,-)$, $(+,-,+)$, $(-,+,+)$, $(-,+,-)$, $(-,-,+)$. These different configurations are associated with different kinds of energy exchange scenario and different stability properties~\cite{W1992}.

We will focus on the two configurations $(-,-,+)$ and $(-,+,+)$, with parameters
\begin{center}
  \begin{enumerate}
  \item $B=-0.0375$, $B_1=-0.025$, $B_2=0.1$
  \item $B=-0.0375$, $B_1=0.025$, $B_2=0.05$
  % \item $B=-0.0375$, $B_1=0.1$, $B_2=-0.025$
  \end{enumerate}
\end{center}
and consider various values of the parameters $\delta$ and $\varepsilon$. The other parameters are fixed to $a=0.01$, $D=0.01$ and $\beta = 0.01/12$. Once the parameterizations have been developed, the different model version have been integrated over $4.5\times 10^{5}$ timeunits with a timestep $\Delta t = 0.01$ after a transient period of $5.0\times 10^4$ timeunits to let the system relax to its stationary state. The state $X$ has been recorded every $0.1$ timeunit, giving a dataset of $4.5\times 10^6$ points for the analysis. The parameterizations given by Eqs.~(\ref{eq:WLres}),~(\ref{eq:majda_params}) and~(\ref{eq:abramov_stoch_def}) have been integrated with a second order Runge-Kutta (RK2) stochastic scheme which converges to the Stratonovich calculus~\cite{HP2006,R1982}. The equation~(\ref{eq:emp_par}) has been integrated with a deterministic RK2 scheme where the stochastic forcing $e(X,t)$ is considered constant during the timestep. The memory term $M_3$ appearing in the parameterization~(\ref{eq:WLres}) and given by the integral~(\ref{eq:M3def}) over the past of $X$ has been computed numerically at each timestep. Although it increases considerably the integration time, this method is adopted in order to clarify the memory effect in Eq.~(\ref{eq:WLres}). A Markovianization of this parameterization is possible~\cite{WDLA2016} but in the present case it would have required some assumptions that would blur the comparison of the methods.

The relative performances of the parameterizations can be tested in multiple ways, by comparing the climatology (the average state) or the variability (variance) of the systems~\cite{N2005}. Another method is to look at the predictive skill score of the models, that is the ability of the parameterizations to provide skillful forecast compared to original system, as in \cite{AMP2013,WDLA2016}. On longer term, the good representation of the ``climate'' of a model by the parameterizations can be assessed by looking at the stationary probability densities and comparing them using some score~\cite{FMV2005,CV2008,A2012,A2013,A2015}. The decorrelation properties of the models and the parameterizations can also be tested, to provide information about the correct representation of the timescales of the models. All those different aspects can be significant, depending on the purpose of the parameterization scheme. However, for the brevity of the present work, we shall focus on the probability densities and whether or not they are correctly reproduced by the parameterizations.

We present now the results obtained the proposed methods and consider first the different measures used for averaging in system~(\ref{eq:triad_sys}).

\subsection{Stability and measures}
\label{sec:stability}

All the ingredients needed to compute the parameterizations presented in Section~\ref{sec:mod_param} can be derived with the help of the covariance and the correlation of the $Y$ variables in the framework of two different systems related to the unresolved dynamics, namely the unperturbed dynamics $\dot Y = F_Y(Y)$ and the unresolved dynamics $\dot Y= H(X,Y)$ with $X$ frozen. The measure of the former is necessary to derive the response theory and the singular perturbation-based parameterizations, while the latter is needed for the Hasselmann averaging method. These two systems are both two-dimensional Ornstein-Uhlenbeck processes of the form
\begin{equation}
  \label{eq:OU}
  \dot Y =  \tens{T}\cdot Y + \tens B \cdot \xi_Y(t)
\end{equation}
for which respectively $\tens T = \tens{A}/\delta^2$ and $\tens T = \tens A/\delta^2 + \left( \varepsilon \, X/\delta\right) \, \tens V \cdot Y$. In both cases, we have $\tens B = \tens B_Y/\delta$. Their measure is then given by~\cite{WDLA2016}
\begin{equation}
  \label{eq:OU_measure}
  \rho(Y) = \frac{1}{\mathcal{Z}} \, \exp\left(- \frac{1}{2} \, Y^\trans \cdot \tens\Sigma^{-1}\cdot Y\right)
\end{equation}
where $\mathcal{Z}$ is a normalization factor and where $\tens\Sigma$ is the covariance matrix solution of
\begin{equation}
  \label{eq:cov}
  \tens T\cdot \tens\Sigma + \tens\Sigma \cdot \tens T^\trans = -\tens B  \cdot \tens B^\trans \quad .
\end{equation}

 In order for theses processes to be stable, the real part of the eigenvalues of the matrix $\tens T$ must be negative~\cite{G2009} for every state $X$ that the full coupled system~(\ref{eq:triad_sys}) can possibly achieve. The eigenvalues of the system $\dot Y = F_Y(Y)$ are $\lambda_\pm=(-a \pm \I \beta)/\delta^2$ and it is thus always stable (since $a>0$ and $\beta \in \mathbb{R}$). On the other hand, the system $\dot Y = H(X,Y)$ has the eigenvalues $\lambda_\pm = (-a \pm \sqrt{\Delta(X))}/\delta^2$ with
\begin{equation}
  \label{eq:delta}
  \Delta(X) = -\left(B_1 X \delta  \epsilon +\beta \right) \left(\beta -B_2 X \delta  \epsilon \right)
\end{equation}
Therefore, if 
\begin{equation}
  \label{eq:condition}
  \Re \left(\sqrt{\Delta(X)}\right)>a 
\end{equation}
for some $X$, the Ornstein-Uhlenbeck process is unstable, and it is then called an \emph{explosive} process. For any initial condition, the process diverges, and thus the only possible stationary measure is the trivial null one. Consequently, equation~(\ref{eq:cov}) gives nonphysical solutions, the stationary covariance matrix does not exist and the parameterizations depending upon cannot be derived.

For the system~(\ref{eq:triad_sys}), if $\mathrm{sgn}(B_1 \, B_2) = -1$,  as in case 1, then the process is stable for every $X$ if $a^2> -\frac{\left(B_1+B_2\right){}^2 \beta ^2}{4 B_1 B_2}$. For case 1, this inequality is satisfied, and thus the process~(\ref{eq:OU}) is stable for every $X$. Moreover, depending on the sign of $\Delta(X)$, the process for $X$ fixed is a stochastic focus (if $\Delta(X)>0$) or a stochastic damped oscillator (if $\Delta(X)<0$). Here, it is a focus if 
\begin{equation}
X\in \left[\min(-\beta/\delta\varepsilon B_1,\beta/\delta\varepsilon B_2),\max(-\beta/\delta\varepsilon B_1,\beta/\delta\varepsilon B_2)\right]  
\end{equation}
and an oscillator otherwise. That is, for the considered $\varepsilon$ and $\delta$ parameters value, the system~(\ref{eq:triad_sys}) is an oscillator for most of the $X$ values.

If $\mathrm{sgn}(B_1 \, B_2) = 1$, as in case 2, then the condition~(\ref{eq:condition}) must be satisfied for every state $X$. For case 2, this inequality was not satisfied for every state $X$ for most of the values of the $\varepsilon$ and $\delta$ parameters considered (see Subsection~\ref{sec:mpp_triad} below). The stability is therefore reversed as the system is non-oscillating for most of the $X$ values.

To summarize, if $B_1$ and $B_2$ are of opposite sign, the dynamics of $\dot Y=H(X,Y)$ is stable and generally oscillatory. If $B_1$ and $B_2$ have the same sign, then the dynamics is unstable in most cases and generally hyperbolic. This is a consequence of the well known difference of stability of the triads depending on their energy exchange properties~\cite{W1992}.

For the interested reader, the exact calculation of the parameterization of Section~\ref{sec:mod_param} using the covariance and correlation matrices are detailed in the appendix (see Section~\ref{sec:appendix}).

\subsection{The $(-,-,+)$ stochastic triad (case 1)}
\label{sec:mmp_triad}
\begin{figure}
  \centering
  \includegraphics[height=0.95\textheight]{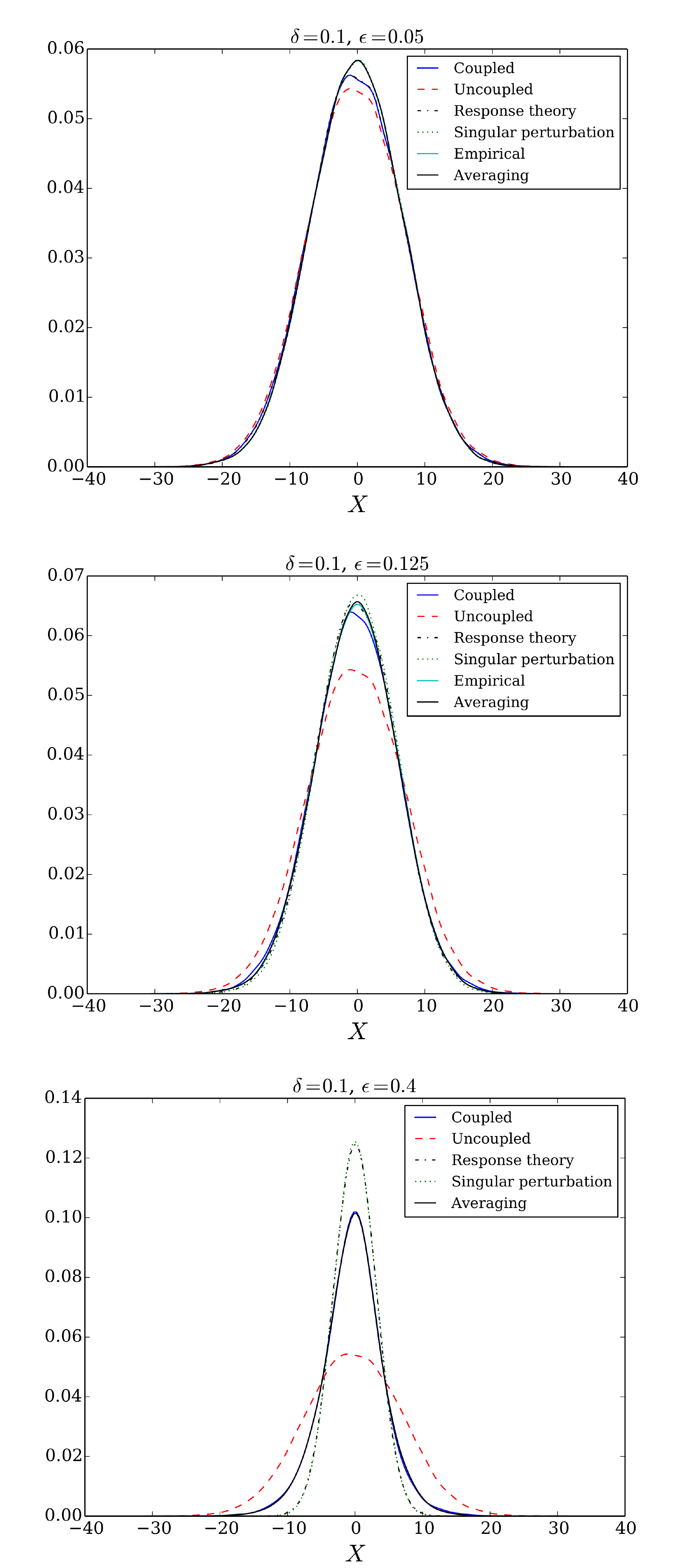}
  \caption{Probability densities of the full coupled dynamics~(\ref{eq:triad_sys}), the uncoupled dynamics $\dot{X} = - D\, X + q\, \xi(t)$ and the parameterized model versions for the timescale separation $\delta=0.1$ and for the triad parameters of case 1. The empirical parameterization density is not represented for $\varepsilon=0.4$ due to its divergence.}
  \label{fig:distrib_case1_delta0.1}
\end{figure}

\begin{figure}
  \centering
  \includegraphics[height=0.95\textheight]{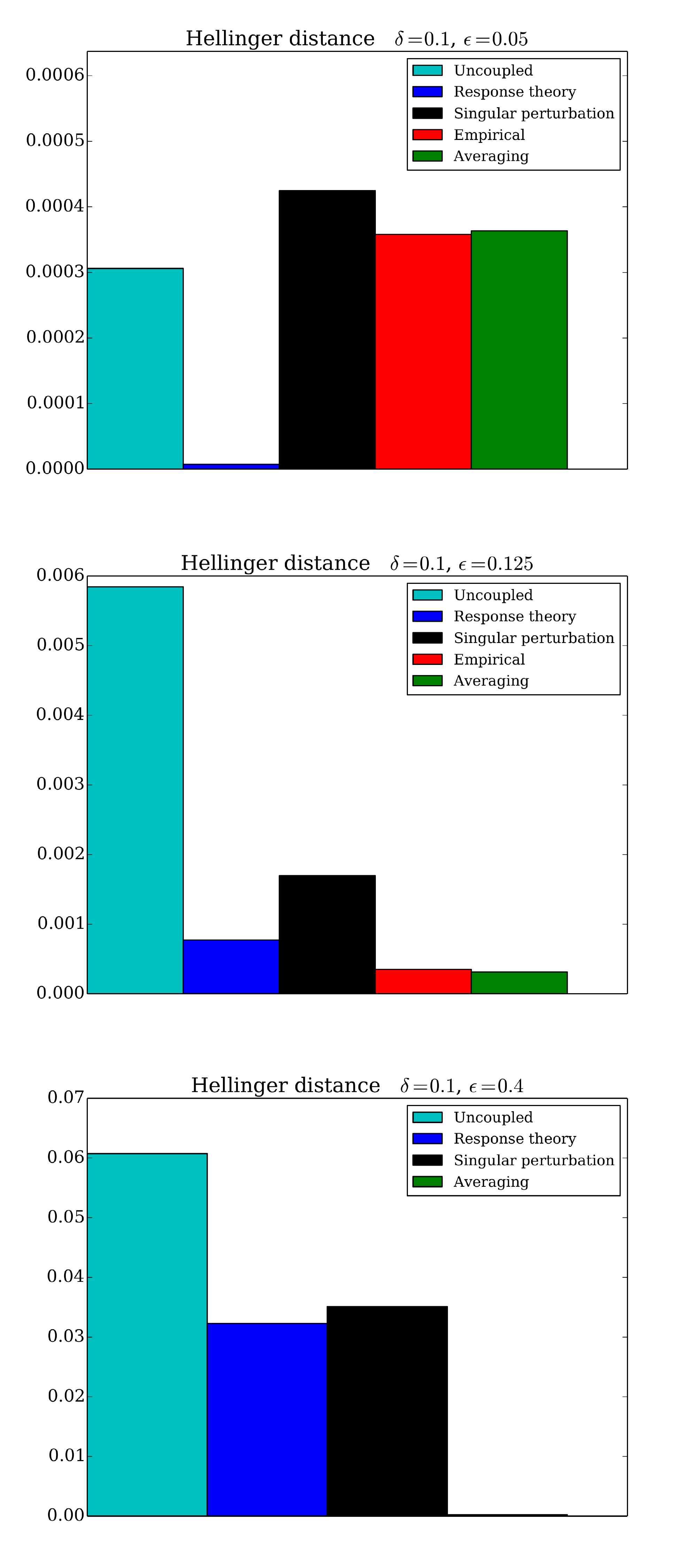}
  \caption{Hellinger distance~(\ref{eq:hellinger}) between the densities of the different parameterized models and the full coupled system density for case 1. A small distance indicates that the two densities concerned are very similar. The Hellinger distance between the full coupled system and the uncoupled system distribution is depicted as reference. In case $\varepsilon=0.4$, the empirical parameterization is not represented due to its divergence.}
  \label{fig:test_case1_delta0.1}
\end{figure}

\begin{figure}
  \centering
  \includegraphics[height=0.95\textheight]{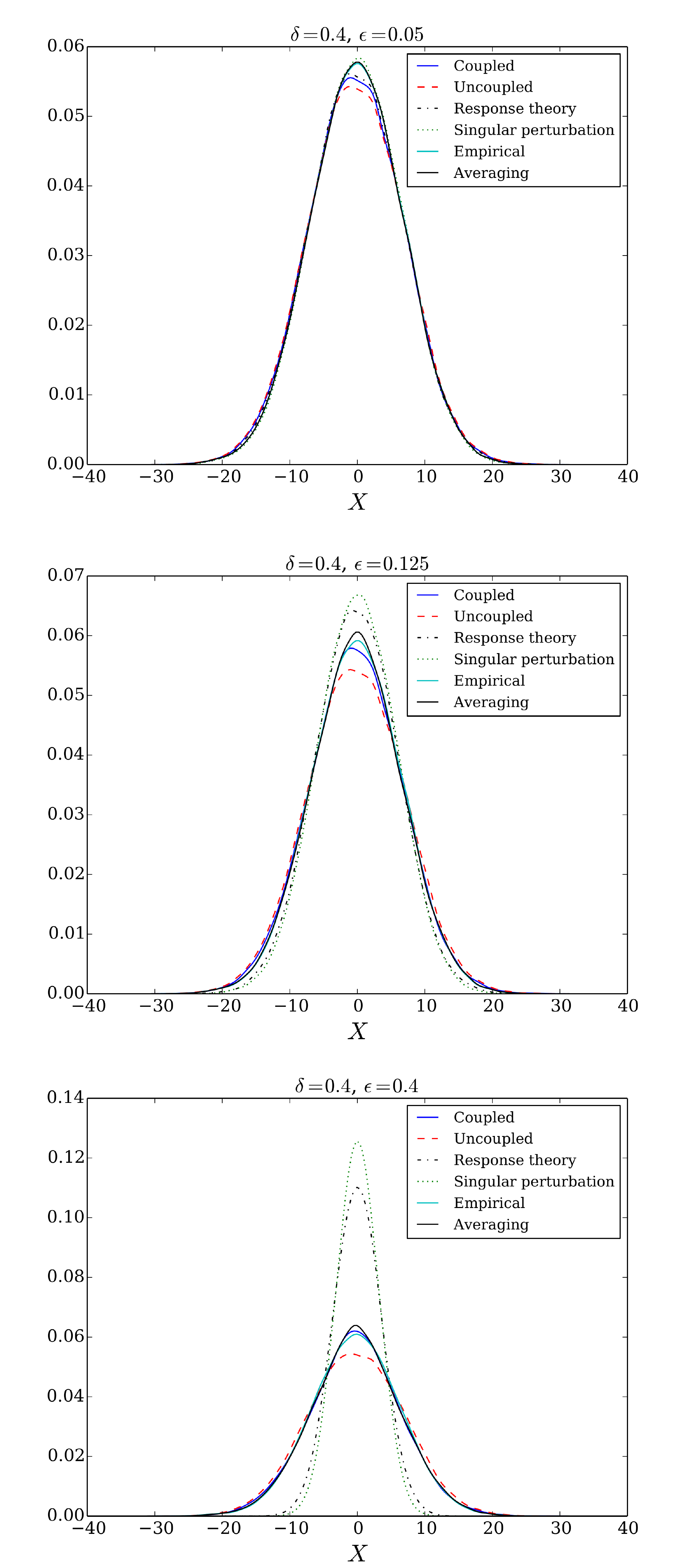}
  \caption{Same as Fig.~\ref{fig:distrib_case1_delta0.1} but for the timescale separation $\delta=0.4$.}
  \label{fig:distrib_case1_delta0.4}
\end{figure}

\begin{figure}
  \centering
  \includegraphics[height=0.95\textheight]{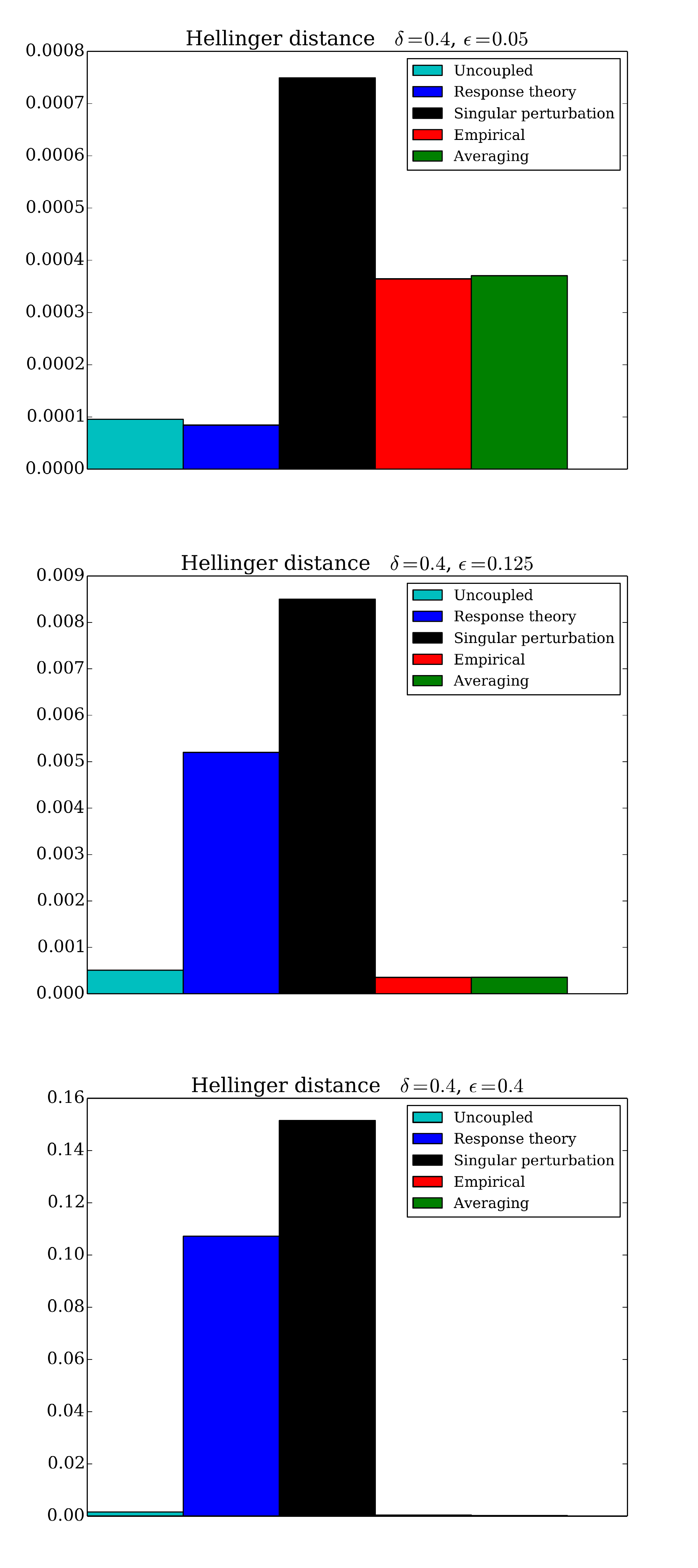}
  \caption{Same as Fig.~\ref{fig:test_case1_delta0.1} but for the timescale separation $\delta=0.4$.}
  \label{fig:test_case1_delta0.4}
\end{figure}

Let us now consider case 1 corresponding to the $(-,-,+)$ stochastic triad for two different values of the timescale separation $\delta=0.1$ and $0.4$. For each of these timescale separation, we considered three values of the coupling strength $\varepsilon$: $0.05$, $0.125$ and $0.4$. The probability densities associated with these different systems are represented on Figs.~\ref{fig:distrib_case1_delta0.1} and~\ref{fig:distrib_case1_delta0.4}. For a timescale separation $\delta=0.1$, the fully coupled dynamics given by Eq.~(\ref{eq:triad_sys}) is quite well represented by all the proposed parameterizations. Since it is hard to distinguish the different density curves, a score such as the Hellinger distance~\cite{AMP2013} 
\begin{equation}
  \label{eq:hellinger}
  H(P,Q) = \frac{1}{2} \int \left( \sqrt{\D P} - \sqrt{\D Q} \right)^2
\end{equation}
between the distribution $P$ of the full coupled system and the distribution $Q$ of the parameterizations is worth computing to quantify the differences. This distance quantifies the similarity between the densities. It is depicted for $\delta=0.1$ on Fig.~(\ref{fig:test_case1_delta0.1}), and it shows that for a very small coupling parameter $\varepsilon=0.05$, the best parameterization is the response theory given by Eq.~(\ref{eq:WLres}). For larger values of $\varepsilon$, it is the Hasselmann averaging method which performs best. The empiric method gives a good correction of the uncoupled dynamics for $\varepsilon=0.125$ but diverges for $\varepsilon=0.4$. This may be due to instabilities introduced by the cubic deterministic parameterization $\mathcal{U}_{\det}(X)$ or to the inadequacy of the fitting function $\sigma_0 + \sigma_1 \, | X |$ for the standard deviation $\sigma_e(X)$ in the AR(1) process~(\ref{eq:emp_ar1}). Indeed, in general, this model fits quite well the statistics in the neighborhood of $X=0$, but the standard deviation reaches a plateau for higher values of $X$. A more complicated fitting function would thus be necessary to get a stable dynamics. 
For an decreased timescale separation $\delta=0.4$, the same conclusions are reached as for $\delta=0.1$, but the singular perturbation method performs not very well in all cases, as illustrated in Fig.~\ref{fig:test_case1_delta0.4} that for $\varepsilon=0.125$ and $0.4$. The response based and singular perturbation methods are even less effective than the uncoupled dynamics. It is not surprising for the latter since it is supposed to be valid in the limit $\delta\to 0$.

\subsection{The $(-,+,+)$ stochastic triad (case 2)}
\label{sec:mpp_triad}

\begin{figure}
  \centering
  \includegraphics[height=0.95\textheight]{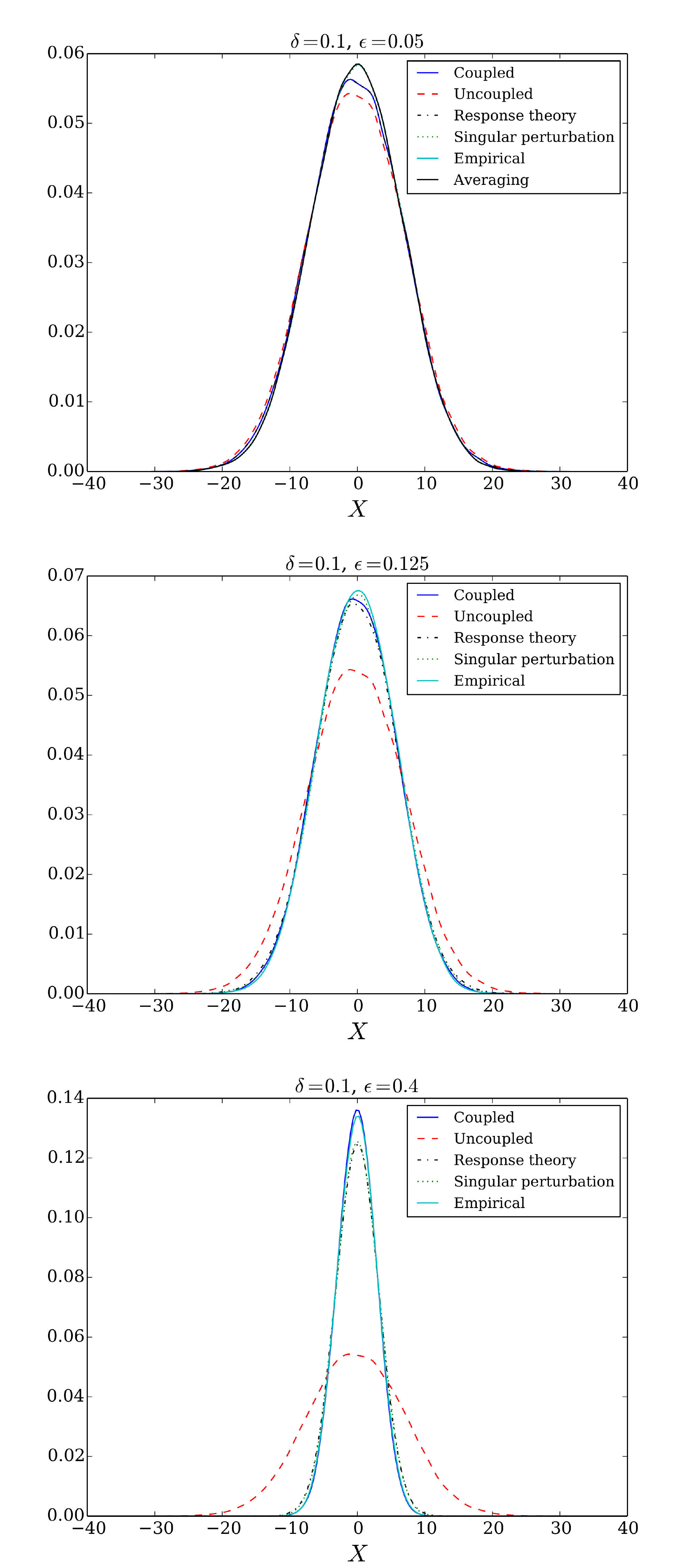}
  \caption{Probability densities of the coupled full dynamics~(\ref{eq:triad_sys}), the uncoupled dynamics $\dot{X} = - D\, X + q\, \xi(t)$ and the parameterized models for the timescale separation $\delta=0.1$ and for the triad parameters of case 2. The direct averaging parameterization density is only represented for $\varepsilon=0.05$ because the system diverges for the other values.}
  \label{fig:distrib_case2_delta0.1}
\end{figure}

\begin{figure}
  \centering
  \includegraphics[height=0.95\textheight]{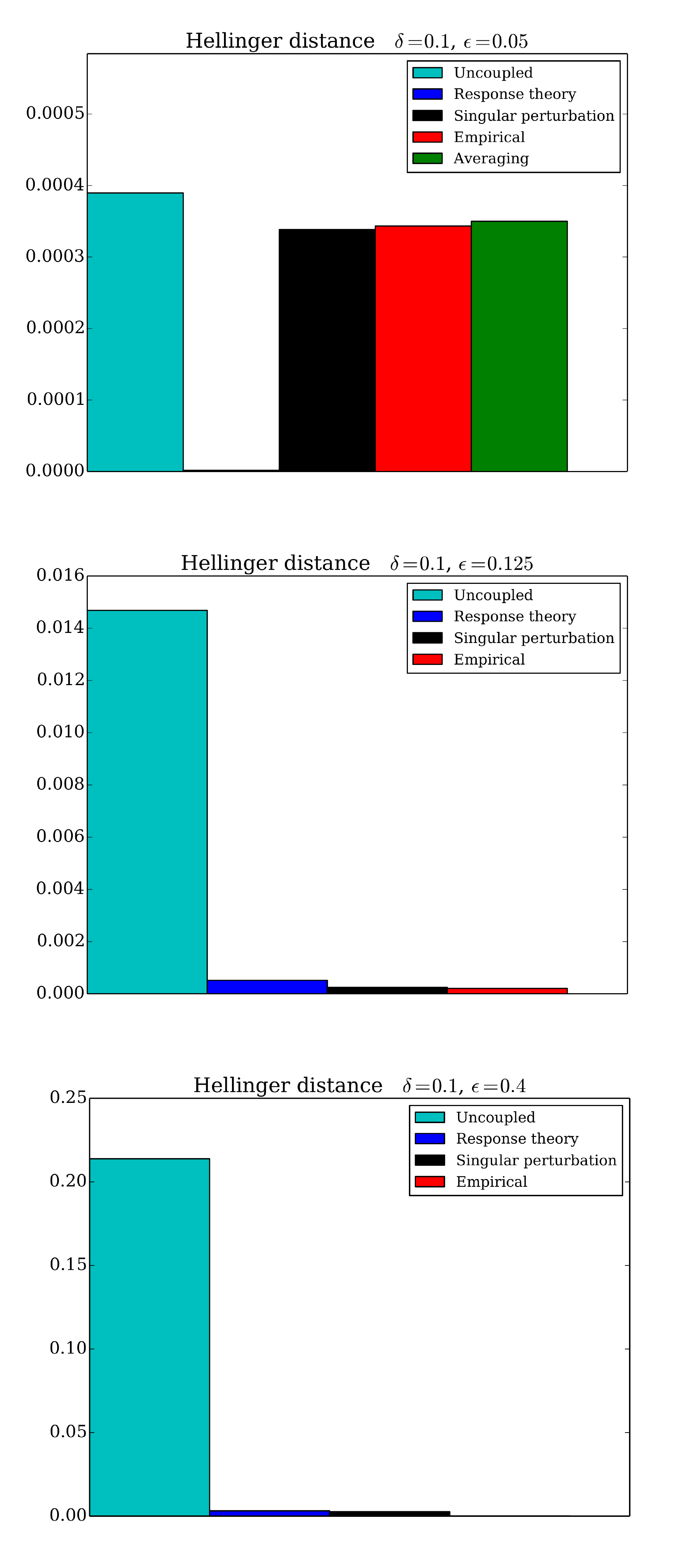}
  \caption{Hellinger distance~(\ref{eq:hellinger}) between the densities of the different parameterized models and the full coupled system density for case 2. A small distance indicates that the two densities concerned are very similar. The Hellinger distance between the full coupled system and the uncoupled system distribution is depicted as reference. In case $\varepsilon=0.4$, the empirical parameterization is not represented due to its divergence.}
  \label{fig:test_case2_delta0.1}
\end{figure}

\begin{figure}
  \centering
  \includegraphics[height=0.95\textheight]{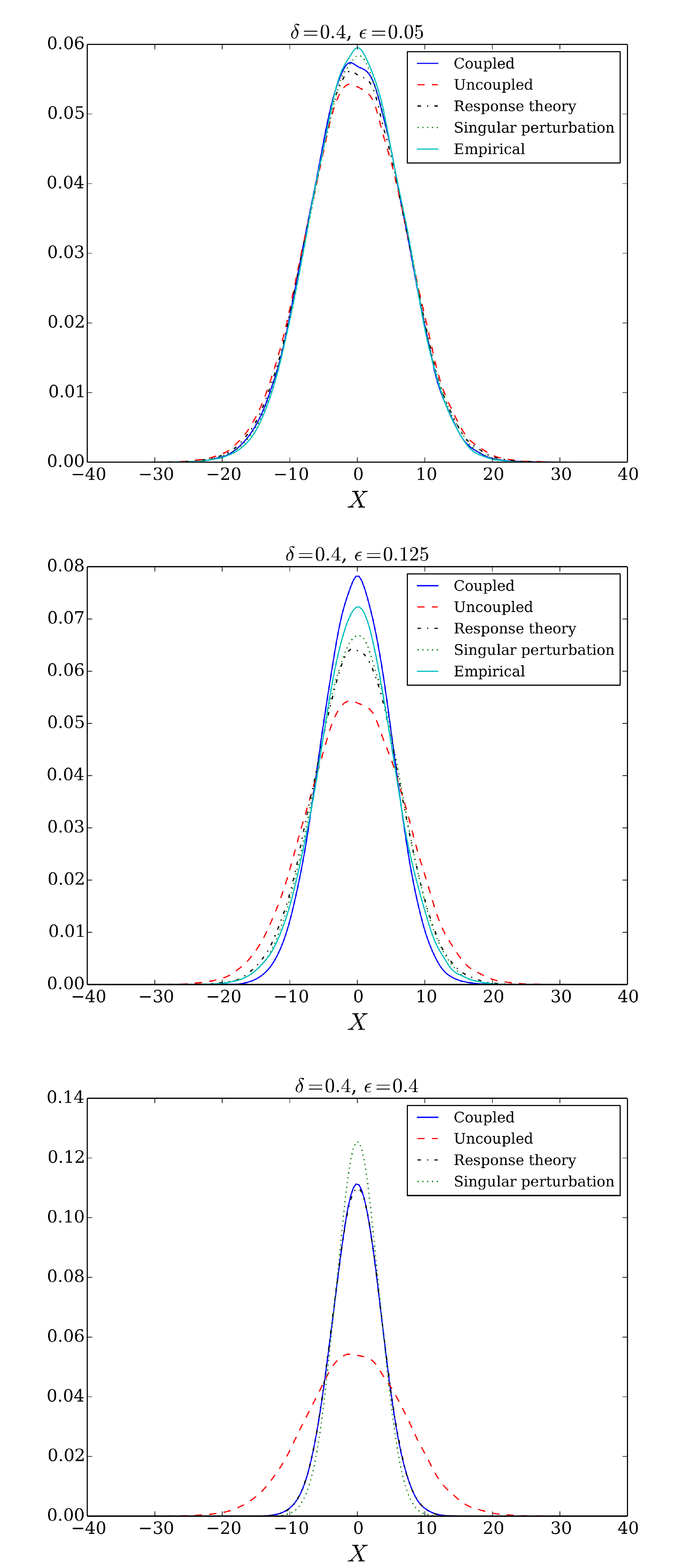}
  \caption{Same as Fig.~\ref{fig:distrib_case2_delta0.1} but for the timescale separation $\delta=0.4$.}
  \label{fig:distrib_case2_delta0.4}
\end{figure}

\begin{figure}
  \centering
  \includegraphics[height=0.95\textheight]{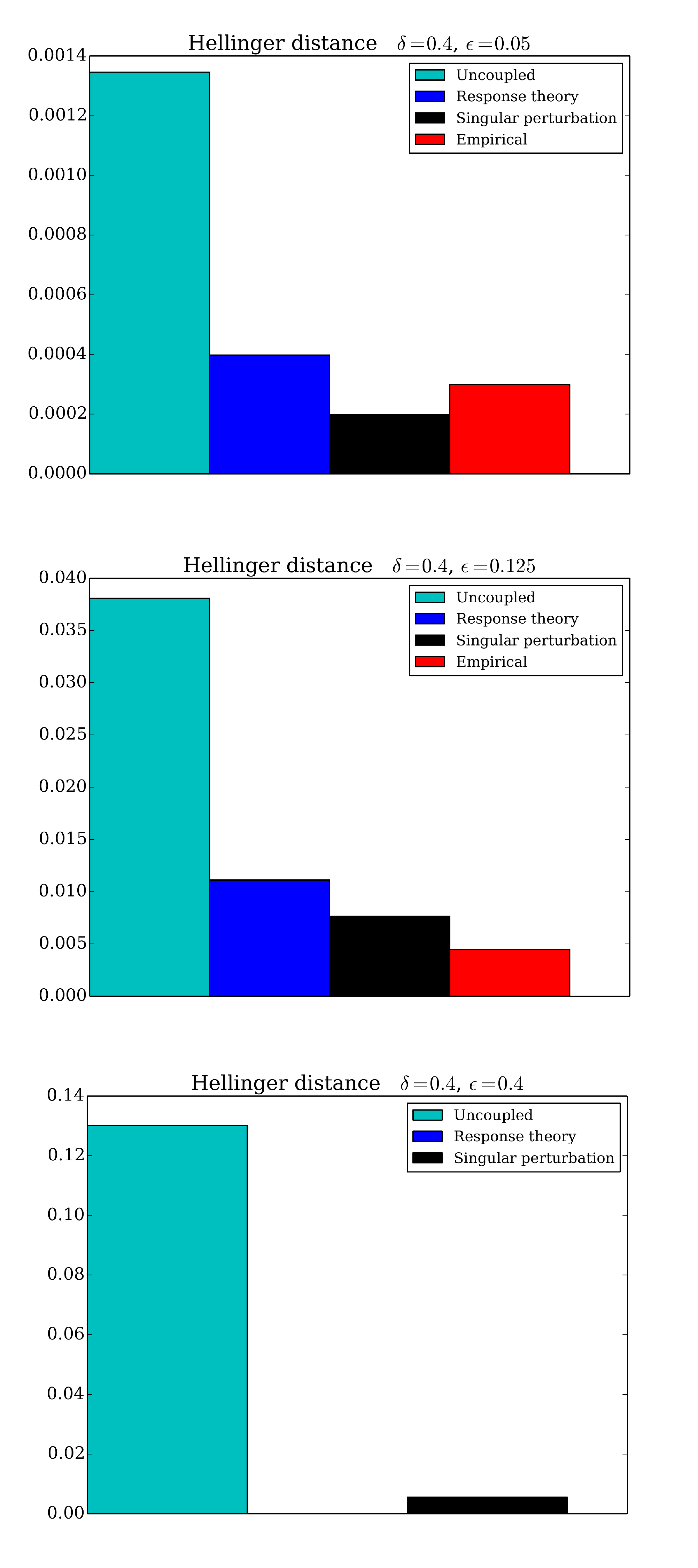}
  \caption{Same as Fig.~\ref{fig:test_case2_delta0.1} but for the timescale separation $\delta=0.4$.}
  \label{fig:test_case2_delta0.4}
\end{figure}

We now consider the parameters of case 2, for which the system~(\ref{eq:triad_sys}) is a $(-,+,+)$ stochastic triad. The probability densities are depicted on Figs.~\ref{fig:distrib_case2_delta0.1} and~\ref{fig:distrib_case2_delta0.4}, and the Hellinger distances are shown on Figs.~\ref{fig:test_case2_delta0.1} and~\ref{fig:test_case2_delta0.4}. First, we must remark that the parameterization based on the Hasselmann's averaging method is not defined for most of the $\delta$ and $\varepsilon$ parameters values. It is due to the fact that the dynamics of the unresolved component $Y$ with $X$ considered as a parameter is unstable, as shown in Subsection~\ref{sec:stability}. Indeed, this linear system undergoes a bifurcation at some value $X^\star$ which destabilizes the dynamics $\dot Y=H(X,Y)$ with $X$ frozen. Therefore, the measure $\rho_{Y|X}$ is not defined for some ranges of the full $X$ dynamics and the method fails. The only case where this destabilization does not occur is for $\delta=0.1$ and $\varepsilon=0.05$, but the parameterization does not perform well. For these parameter values, the only parameterization that performs very well is the one based on response theory. For the other values of the parameters $\delta$ and $\varepsilon$, all the parameterizations have good performances. A particularly unexpected result is the very good correction provided by the response theory and singular perturbation based methods for the extreme case $\delta=0.4$ and $\varepsilon=0.4$ (see the bottom panel of Fig.~\ref{fig:test_case2_delta0.4}). This have to be contrasted with their bad performances in the case of the other triad (see the bottom panel of Fig.~\ref{fig:test_case1_delta0.4}). Note that for this extreme case, the direct averaging method fails and the empirical method is unstable and diverges.

\subsection{Discussion}
\label{sec:discu}

The results obtained so far with these two types of triads highlight the utility of the parameterization schemes discussed here. First, the empiric parameterization gives usually good results when it does not destabilize the dynamics. However, this method requires a case by case time-consuming statistical analysis whose complexity increases with the dimensionality of the problem considered. Physically based parameterizations do not require such an analysis, and the best approach in the present system is the Hasselmann averaging one, but it requires that the dynamics of the unresolved system be stable. It was thus very effective to correct the dynamics of the $(-,-,+)$ triad, but not the other triad $(-,+,+)$. In this latter case, the perturbative methods like the singular perturbation method or the response theory method gives very good results. This difference is quite intriguing and interesting. It indicates that different physically based parameterization should be considered depending on the kind of problems encountered. In particular, the stability properties of the system considered seems to play an important role. This conclusion holds whatever the timescale separation and for the most realistic values of the coupling strength between the components ($\varepsilon=0.125$ and $0.4$). However, for very small values of the coupling strength, the response based method seems to be the best approach in all cases.

A question that is left open in the present work is to determine precisely which stability property is giving the contrasting observed result. In extenso, is it the hyperbolic instability of the $(-,+,+)$ triad which makes the perturbative approach and the response based parameterization perform so well, or is it simply the fact that it is unstable? On the other hand, is it the damped oscillatory behavior of the $(-,-,+)$ triad which makes the Hasselmann's method works well, or is it simply the fact that it is stable? Such questions should be addressed in the case of a more complex, globally stable system, which allows to have locally stable and unstable fast dynamics.

\section{Conclusions}
\label{sec:conclu}

The parameterization of subgrid-scale processes is an important tool available for meteorologist and climatologist, for stochastic modeling, model reduction, or to improve the statistical properties of the forecasting systems. The variety of different approaches available bear witness of the richness of the field but at the same time can also lead to questions on the best choice for the problem at hand. The purpose of the present review was to describe briefly some of the most recent methods and to illustrate them on a simple stochastic triad example. The methods covered included perturbative methods like the Ruelle response theory~\cite{WL2012} or the singular perturbation theory~\cite{MTV2001}, averaging methods like the Hasselmann method~\cite{H1976,AIW2003} or empirical methods~\cite{AMP2013}. As expected, these parameterizations provided good results depending on the timescale separation and on the coupling between the resolved variables and the subgrid one. But more importantly, our results in the context of this simple triad stress the importance of the underlying stability properties of the unresolved system. It thus confirms a known result that the structure of the Jacobian and of the Hessian of a given system control the behavior and performance of model error parameterizations~\cite{N2005}. 

Further comparisons of the different methods are needed in the context of more sophisticated systems in order to analyze the role of the stability properties of the sub-grid scale processes on their performances. This type of analysis is currently under way in the context of a coupled ocean-atmosphere system~\cite{DDV2016}.
\appendix

\section{Appendix : Practical computation of the parameterizations}
\label{sec:appendix}

In the following section, for illustrative purposes, we detail the computation that we have made to obtain the result of the present review. We start with the method based on response theory.
\subsection{Response theory method}
\label{sec:app_WL}
We consider the system~(\ref{eq:triad_sys}) with the form~(\ref{eq:dec_gen_sys_spec}) in mind. In this case, the influence of the $Y$ sub-system on the $X$ sub-system is parameterized as:
\begin{equation}
  \label{eq:app_WL_params}
  \dot X = - D\, X + q \, \xi(t) + M_1(X) + M_2(X,t) +M_3(X,t)
\end{equation}
where then terms $M_1$,$M_2$ and $M_3$ are respectively given by Eqs.~(\ref{eq:M1def}),~(\ref{eq:gdef} and~(\ref{eq:M3def}). The average in these formula are performed with the measure $\rho_{0,Y}$ of the unperturbed $Y$ dynamics $\dot Y=F_Y(Y)$. Since this latter is an Ornstein-Uhlenbeck process, its measure is the Wiener measure
\begin{equation}
  \label{eq:app_Wiener}
  \rho_{0,Y}(Y) = \frac{1}{\mathcal{Z}} \, \exp\left(- \frac{1}{2} \, Y^\trans \cdot \tens\Sigma^{-1}\cdot Y\right)
\end{equation}
where $\tens\Sigma$ is the covariance matrix solution of
\begin{equation}
  \label{eq:app_covA}
  \tens A\cdot \Sigma + \Sigma \cdot \tens A^\trans = - \tens B_Y  \cdot \tens B_Y^\trans \quad
\end{equation}
and $\mathcal{Z}$ is a normalization factor.

The covariance and correlation of the stationary process $\dot Y=F_Y(Y)$ are thus straightforward to compute~\cite{G2009}:
\begin{eqnarray}
  \label{eq:app_Y_corr_cov}
  \tens\Sigma & = & \big\langle Y \otimes Y \big\rangle = \frac{q^2_Y}{2 a} \, \tens I \\
    \big\langle \phi_Y^t(Y) \otimes \phi_Y^s(Y) \big\rangle & = & \tens E(t-s) \cdot \tens\Sigma \quad , \quad t > s \\
    \big\langle \phi_Y^t(Y) \otimes \phi_Y^s(Y) \big\rangle & = & \tens\Sigma \cdot \tens E(s-t)^\trans \quad , \quad t < s
\end{eqnarray}
where $\tens I$ is the identity matrix, $\phi_Y^t$ is the flow of $\dot Y = F_Y(Y)$ and the matrix $\tens E(t)$ is the exponential
\begin{equation}
  \label{eq:app_mat_exp}
  \tens E(t) = \exp(\tens A t/\delta^2) = e^{-a t/\delta^2} \left[
    \begin{array}{cc}
      \cos(\beta t/\delta^2) & \sin(\beta t/\delta^2) \\ -\sin(\beta t/\delta^2) & \cos(\beta t/\delta^2)
    \end{array}
    \right]
\end{equation}
The various terms $M_i$ are then computed as follow.
\subsubsection{The term $M_1$}
\label{sec:M1}

It is the average term:
\begin{equation}
  \label{eq:app_M1_def}
  M_1(X) = \big\langle \Psi_X(X,Y) \big\rangle
\end{equation}
We have thus
\begin{equation}
  \label{eq:app_M1_res}
  M_1(X) = 2 B \frac{\varepsilon}{\delta} \langle y_1 \, y_2 \rangle = 2 B \frac{\varepsilon}{\delta} \, \Sigma_{12} = 0
\end{equation}
by using Eq.~(\ref{eq:app_Y_corr_cov}).

\subsubsection{The term $M_2$}
\label{sec:M2}

It is the noise/correlation term which is defined here as:
\begin{equation}
  \label{eq:app_M2_def}
  M_2(t) = \sigma_R(t)
\end{equation}
with
\begin{equation}
  \label{eq:app_M2_corr}
  \langle \sigma_R(t) \sigma_R(t')\rangle = g(t-t')
\end{equation}
and the correlation function
\begin{equation}
  \label{eq:app_g_def}
  g (s) = \big\langle \Psi_X^\prime (Y) \Psi_X^\prime(\phi_Y^s(Y))\big\rangle
\end{equation}
where $\Psi_X^\prime (Y) = \Psi_X (Y) - M_1$.
The result in the present case is given by the formula (see Ref.~\cite{DV2016}) :
\begin{equation}
  \label{eq:app_M2_formula}
  \tens g(s) = \frac{\varepsilon^2}{\delta^2} \mathrm{Tr} \left( \left( \tens C + \tens C^\trans\right) \cdot \tens \Sigma \cdot \tens E(s)^\trans \cdot \tens C^\trans \cdot \tens E(s) \cdot \tens \Sigma \right) = \frac{\varepsilon^2}{\delta^2} \frac{q_Y^4}{a^2} B^2 e^{-2 a s/\delta^2} \cos(2 \beta s/\delta^2)
\end{equation}
The term $M_2$ must thus be devised as a process with the same correlation.
\subsubsection{The term $M_3$}
\label{sec:M3}

This is the memory term, defined by
\begin{equation}
  \label{eq:app_M3_def}
  M_3(X,t) = \int_0^\infty \, \D s \,\, h(X(t-s),s)
\end{equation}
with the memory kernel
\begin{equation}
  \label{eq:app_mem_ker}
  h(X,s) = \left\langle \Psi_Y(X,Y) \cdot \nabla_Y \Psi_X(Y^s)\right\rangle
\end{equation}
which in the present case is given by the formula (see Ref.~\cite{DV2016})
\begin{eqnarray}
  h(X,s) & = & \frac{\varepsilon^2}{\delta^2} \mathrm{Tr} \left(\left(X \, \tens V \cdot \tens \Sigma\right) \cdot \left(\tens E(s)^\trans \cdot \left(\tens C + \tens C^\trans\right) \cdot \tens E(s)\right)\right) \\ 
  & = & \frac{\varepsilon^2}{\delta^2} \frac{q_Y^2}{a} X B (B_1 +B_2) e^{-2 a s/\delta^2} \cos(2 \beta s/\delta^2)   \label{eq:app_mem_ker_formula}
\end{eqnarray}
The fact that the memory kernel~(\ref{eq:app_mem_ker_formula}) and the correlation function~(\ref{eq:app_M2_formula}) present the same form imply that a Markovian parameterization is available~\cite{WDLA2016} even if by definition, Eq.~(\ref{eq:app_WL_params}) is a non-Markovian parameterization.

\subsection{The singular perturbation method}
\label{sec:app_majda}
With this parameterization, the parameter $\delta$ serves to distinguish terms with different timescale and is then used as a small perturbation parameter~\cite{MTV2001,FMV2005}.
The parameterization is given by:
\begin{equation}
  \label{eq:app_majda_params}
  \dot X = - D\, X + q\, \xi(t) +G(X) + \sqrt{2}\, \sigma_{\rm MTV}(X) \cdot \tilde\xi(t)
\end{equation}
with notably $\langle \xi(t) \tilde\xi(t') \rangle =0$ and
\begin{eqnarray}
  \label{eq:app_majda_def}
  G(X) &= & \int_0^\infty \, \D s \, \big\langle \Psi_Y(X,Y)\cdot \nabla_Y \Psi_X(X,\phi^s_Y(Y))\big\rangle_{\tilde\rho} \\
  \sigma_{\rm MTV}(X) & = & \left(\int_0^\infty \, \D s \, \big\langle \Psi_X^\prime(X,Y) \, \Psi_X^\prime(X,\phi^s_Y(Y)) \big\rangle_{\tilde\rho}\right)^{1/2} \label{eq:app_SMTVdef}
\end{eqnarray}
We see that the quantities appearing in this parameterization can easily be obtained from the functions $h$ and $g$ of the section~\ref{sec:app_WL}. Indeed we have,
\begin{eqnarray}
  G(X) & = & \int_0^\infty d\! s \,\, h(X,s) = \varepsilon^2 X q_Y^2 \frac{ B (B_1+B_2)}{2(a^2+\beta^2)} \\
  \tens S_{\rm MTV}(X) & = & \int_0^\infty d\! s \,\, g(s) = \varepsilon^2 \frac{q_Y^4 B^2}{2 a (a^2+\beta^2)}
\end{eqnarray}
where we notice that the parameter $\delta$ has disappeared, since this parameterization is valid in the limit $\delta\to 0$.

\subsection{Averaging method}
\label{sec:app_abramov_det}
In this approach, we consider the system~(\ref{eq:dec_gen_sys}) and the parameterization from Ref.~\cite{A2013}:
\begin{equation}
  \label{eq:app_abramoc_det_params}
  \dot X = \bar F (X)
\end{equation}
with
\begin{equation}
  \label{eq:app_averaged_F}
  \bar F(X) = \big\langle F (X,Y) \big\rangle_{\rho_{Y|X}} = F(X,\bar Y(X)) + \frac{1}{2} \frac{\partial^2 F}{\partial Y^2} (X,\bar Y(X)) : \Sigma(X)
\end{equation}
and
\begin{eqnarray}
  \label{eq:app_zdef}
   \bar Y(X) & = & \langle Y \rangle_{\rho_{Y|X}} \\
  \tens \Sigma(X) & = & \langle (Y-\bar Y(X) \otimes (Y-\bar Y(X)) \rangle_{\rho_{Y|X}}
\end{eqnarray}
where $\rho_{Y|X}$ is the measure of the system $\dot Y=H(X,Y)$ with $X$ ``frozen''. It is the measure of an Ornstein-Uhlenbeck process
\begin{equation}
  \label{eq:app_WienerX}
  \rho_{0,Y}(Y) = \frac{1}{\mathcal{Z}} \, \exp\left(- \frac{1}{2} \, Y^\trans \cdot \tens\Sigma^{-1}(X)\cdot Y\right)
\end{equation}
where $\mathcal{Z}$ is a normalization factor and $\tens\Sigma(X)$ is the stationary covariance matrix solution of
\begin{equation}
  \label{eq:app_covT}
  \tens T(X) \cdot \tens\Sigma + \tens\Sigma \cdot \tens T(X)^\trans = - \frac{1}{\delta^2} \tens B_Y  \cdot \tens B_Y^\trans
\end{equation}
with
\begin{equation}
  \label{eq:app_T_def}
  \tens T(X) = \tens A/\delta^2 + \varepsilon \, X \, \tens V/\delta \quad .
\end{equation}
With the help of $\bar Y(X)=0$ and $\tens\Sigma(X)$, we can now rewrite Eq.~(\ref{eq:app_averaged_F}) as
\begin{eqnarray}
  \label{eq:app_avF}
  \bar F (X) & = & F(X,0) + \frac{\varepsilon}{\delta}\, \tens C : \tens\Sigma(X) \nonumber \\
  & = & - D \, X + q \xi(t) + \frac{B \left(B_1+B_2\right)  q_Y^2 X \epsilon ^2}{2\left(a^2+\beta ^2-X \beta  \delta  \varepsilon  B_2+X \delta \varepsilon  B_1 \left(\beta -X \delta  \varepsilon  B_2\right)\right)}
\end{eqnarray}
This forms a deterministic averaging parameterization. It can be extended into a stochastic parameterization~\cite{A2015} as follow:
\begin{equation}
  \label{eq:app_abramov_stoch_def}
  \dot X = \bar F (X) + \sigma_{\rm A} (X) \cdot \xi(t)
\end{equation}
with 
\begin{equation}
  \label{eq:app_abramov_sigma}
  \sigma_{\rm A} (X) = \sqrt{ S(X)}
\end{equation}
and
\begin{equation}
  \label{eq:app_S_def}
  S(X)=2 \int_0^\infty \, \D s \, \, \left\langle \big(F(X,Y^s(X))-\bar F(X)\big)\otimes\big(F(X,Y)-\bar F(X)\big)\right\rangle_{\rho_{Y|X}}
\end{equation}
We thus have
\begin{eqnarray}
  \label{eq:app_S_sol}
  S(X) & = & 2 \frac{\varepsilon^2}{\delta^2} \int_0^\infty \, \D s \,\, \mathrm{Tr} \left(\left( \tens C + \tens C^\trans\right) \cdot \tens\Sigma(X) \cdot \exp\left[\tens T(X)^\trans s\right] \cdot \tens C^\trans \cdot \exp\left[\tens T(X) s\right] \cdot \tens \Sigma(X) \right) \nonumber % \\
  % & = & \frac{1}{8 a \left(a^2+\beta
  %     ^2+\delta  X \epsilon  \left(\beta  B_1-B_2 \left(B_1 \delta  X \epsilon +\beta \right)\right)\right){}^3} \\
  % & & \times B^2 \delta ^2 q_Y^4 \left(4 \left(a^2+\beta ^2\right)^2+\delta  X \epsilon  \left(B_2^2 \delta  X \epsilon  \left(5
  %       \left(a^2+\beta ^2\right) \\
  %       & & +B_1 \delta  X \epsilon  \left(2 B_1 \delta  X \epsilon +7 \beta \right)\right)-B_2 \left(8
  %       \beta  \left(a^2+\beta ^2\right) \\
  %       & & +B_1 \delta  X \epsilon  \left(B_1 \delta  X \epsilon  \left(B_1 \delta  X \epsilon
  %  +7 \beta \right)-2 \left(a^2-7 \beta ^2\right)\right)\right)+B_1 \left(8 \beta  \left(a^2+\beta ^2\right)+B_1 \delta
  %   X \epsilon  \left(5 \left(a^2+\beta ^2\right)+\beta  B_1 \delta  X \epsilon \right)\right)+B_2^3 \delta ^2
  %  \left(-X^2\right) \epsilon ^2 \left(B_1 \delta  X \epsilon +\beta \right)\right)\right)
\end{eqnarray}
where we have extended the result of Eq.~(\ref{eq:app_M2_formula}) to the stationary Ornstein-Uhlenbeck process $\dot Y=H(X,Y)$ for X ``frozen''. The function $S(X)$ can be computed analytically using mathematical softwares but is a very complicate function that is not worth displaying in this review. This can however be provided upon query to Dr. J. Demaeyer.

\bibliography{art_irm}

\begin{thebibliography}{65}%
\makeatletter
\providecommand \@ifxundefined [1]{%
 \@ifx{#1\undefined}
}%
\providecommand \@ifnum [1]{%
 \ifnum #1\expandafter \@firstoftwo
 \else \expandafter \@secondoftwo
 \fi
}%
\providecommand \@ifx [1]{%
 \ifx #1\expandafter \@firstoftwo
 \else \expandafter \@secondoftwo
 \fi
}%
\providecommand \natexlab [1]{#1}%
\providecommand \enquote  [1]{``#1''}%
\providecommand \bibnamefont  [1]{#1}%
\providecommand \bibfnamefont [1]{#1}%
\providecommand \citenamefont [1]{#1}%
\providecommand \href@noop [0]{\@secondoftwo}%
\providecommand \href [0]{\begingroup \@sanitize@url \@href}%
\providecommand \@href[1]{\@@startlink{#1}\@@href}%
\providecommand \@@href[1]{\endgroup#1\@@endlink}%
\providecommand \@sanitize@url [0]{\catcode `\\12\catcode `\$12\catcode
  `\&12\catcode `\#12\catcode `\^12\catcode `\_12\catcode `\%12\relax}%
\providecommand \@@startlink[1]{}%
\providecommand \@@endlink[0]{}%
\providecommand \url  [0]{\begingroup\@sanitize@url \@url }%
\providecommand \@url [1]{\endgroup\@href {#1}{\urlprefix }}%
\providecommand \urlprefix  [0]{URL }%
\providecommand \Eprint [0]{\href }%
\providecommand \doibase [0]{http://dx.doi.org/}%
\providecommand \selectlanguage [0]{\@gobble}%
\providecommand \bibinfo  [0]{\@secondoftwo}%
\providecommand \bibfield  [0]{\@secondoftwo}%
\providecommand \translation [1]{[#1]}%
\providecommand \BibitemOpen [0]{}%
\providecommand \bibitemStop [0]{}%
\providecommand \bibitemNoStop [0]{.\EOS\space}%
\providecommand \EOS [0]{\spacefactor3000\relax}%
\providecommand \BibitemShut  [1]{\csname bibitem#1\endcsname}%
\let\auto@bib@innerbib\@empty
%</preamble>
\bibitem [{\citenamefont {Olbers}(2001)}]{O2001}%
  \BibitemOpen
  \bibfield  {author} {\bibinfo {author} {\bibfnamefont {D.}~\bibnamefont
  {Olbers}},\ }in\ \href@noop {} {\emph {\bibinfo {booktitle} {Stochastic
  Climate Models}}}\ (\bibinfo  {publisher} {Springer},\ \bibinfo {year}
  {2001})\ pp.\ \bibinfo {pages} {3--63}\BibitemShut {NoStop}%
\bibitem [{\citenamefont {Nicolis}\ and\ \citenamefont
  {Nicolis}(1981)}]{NN1981}%
  \BibitemOpen
  \bibfield  {author} {\bibinfo {author} {\bibfnamefont {C.}~\bibnamefont
  {Nicolis}}\ and\ \bibinfo {author} {\bibfnamefont {G.}~\bibnamefont
  {Nicolis}},\ }\href@noop {} {\bibfield  {journal} {\bibinfo  {journal}
  {Tellus}\ }\textbf {\bibinfo {volume} {33}},\ \bibinfo {pages} {225}
  (\bibinfo {year} {1981})}\BibitemShut {NoStop}%
\bibitem [{\citenamefont {Nicolis}\ and\ \citenamefont
  {Nicolis}(2012)}]{NN2012}%
  \BibitemOpen
  \bibfield  {author} {\bibinfo {author} {\bibfnamefont {G.}~\bibnamefont
  {Nicolis}}\ and\ \bibinfo {author} {\bibfnamefont {C.}~\bibnamefont
  {Nicolis}},\ }\href@noop {} {\emph {\bibinfo {title} {Foundations of complex
  systems: emergence, information and predicition}}}\ (\bibinfo  {publisher}
  {World Scientific},\ \bibinfo {year} {2012})\BibitemShut {NoStop}%
\bibitem [{\citenamefont {Hasselmann}(1976)}]{H1976}%
  \BibitemOpen
  \bibfield  {author} {\bibinfo {author} {\bibfnamefont {K.}~\bibnamefont
  {Hasselmann}},\ }\href@noop {} {\bibfield  {journal} {\bibinfo  {journal}
  {Tellus A}\ }\textbf {\bibinfo {volume} {28}} (\bibinfo {year}
  {1976})}\BibitemShut {NoStop}%
\bibitem [{\citenamefont {Ghil}\ \emph {et~al.}(2002)\citenamefont {Ghil},
  \citenamefont {Allen}, \citenamefont {Dettinger}, \citenamefont {Ide},
  \citenamefont {Kondrashov}, \citenamefont {Mann}, \citenamefont {Robertson},
  \citenamefont {Saunders}, \citenamefont {Tian}, \citenamefont {Varadi} \emph
  {et~al.}}]{Ga2002}%
  \BibitemOpen
  \bibfield  {author} {\bibinfo {author} {\bibfnamefont {M.}~\bibnamefont
  {Ghil}}, \bibinfo {author} {\bibfnamefont {M.}~\bibnamefont {Allen}},
  \bibinfo {author} {\bibfnamefont {M.}~\bibnamefont {Dettinger}}, \bibinfo
  {author} {\bibfnamefont {K.}~\bibnamefont {Ide}}, \bibinfo {author}
  {\bibfnamefont {D.}~\bibnamefont {Kondrashov}}, \bibinfo {author}
  {\bibfnamefont {M.}~\bibnamefont {Mann}}, \bibinfo {author} {\bibfnamefont
  {A.~W.}\ \bibnamefont {Robertson}}, \bibinfo {author} {\bibfnamefont
  {A.}~\bibnamefont {Saunders}}, \bibinfo {author} {\bibfnamefont
  {Y.}~\bibnamefont {Tian}}, \bibinfo {author} {\bibfnamefont {F.}~\bibnamefont
  {Varadi}},  \emph {et~al.},\ }\href@noop {} {\bibfield  {journal} {\bibinfo
  {journal} {Reviews of geophysics}\ }\textbf {\bibinfo {volume} {40}}
  (\bibinfo {year} {2002})}\BibitemShut {NoStop}%
\bibitem [{\citenamefont {Lovejoy}\ and\ \citenamefont
  {Schertzer}(2013)}]{LS2013}%
  \BibitemOpen
  \bibfield  {author} {\bibinfo {author} {\bibfnamefont {S.}~\bibnamefont
  {Lovejoy}}\ and\ \bibinfo {author} {\bibfnamefont {D.}~\bibnamefont
  {Schertzer}},\ }\href@noop {} {\emph {\bibinfo {title} {The weather and
  climate: emergent laws and multifractal cascades}}}\ (\bibinfo  {publisher}
  {Cambridge University Press},\ \bibinfo {year} {2013})\BibitemShut {NoStop}%
\bibitem [{\citenamefont {Lemke}(1977)}]{L1977}%
  \BibitemOpen
  \bibfield  {author} {\bibinfo {author} {\bibfnamefont {P.}~\bibnamefont
  {Lemke}},\ }\href@noop {} {\bibfield  {journal} {\bibinfo  {journal}
  {Tellus}\ }\textbf {\bibinfo {volume} {29}},\ \bibinfo {pages} {385}
  (\bibinfo {year} {1977})}\BibitemShut {NoStop}%
\bibitem [{\citenamefont {Frankignoul}\ and\ \citenamefont
  {Hasselmann}(1977)}]{FH1977}%
  \BibitemOpen
  \bibfield  {author} {\bibinfo {author} {\bibfnamefont {C.}~\bibnamefont
  {Frankignoul}}\ and\ \bibinfo {author} {\bibfnamefont {K.}~\bibnamefont
  {Hasselmann}},\ }\href@noop {} {\bibfield  {journal} {\bibinfo  {journal}
  {Tellus}\ }\textbf {\bibinfo {volume} {29}},\ \bibinfo {pages} {289}
  (\bibinfo {year} {1977})}\BibitemShut {NoStop}%
\bibitem [{\citenamefont {Frankignoul}(1979)}]{F1979}%
  \BibitemOpen
  \bibfield  {author} {\bibinfo {author} {\bibfnamefont {C.}~\bibnamefont
  {Frankignoul}},\ }\href@noop {} {\bibfield  {journal} {\bibinfo  {journal}
  {Dynamics of Atmospheres and Oceans}\ }\textbf {\bibinfo {volume} {3}},\
  \bibinfo {pages} {465} (\bibinfo {year} {1979})}\BibitemShut {NoStop}%
\bibitem [{\citenamefont {Frankignoul}\ and\ \citenamefont
  {M{\"u}ller}(1979)}]{FM1979}%
  \BibitemOpen
  \bibfield  {author} {\bibinfo {author} {\bibfnamefont {C.}~\bibnamefont
  {Frankignoul}}\ and\ \bibinfo {author} {\bibfnamefont {P.}~\bibnamefont
  {M{\"u}ller}},\ }\href@noop {} {\bibfield  {journal} {\bibinfo  {journal}
  {Journal of Physical Oceanography}\ }\textbf {\bibinfo {volume} {9}},\
  \bibinfo {pages} {104} (\bibinfo {year} {1979})}\BibitemShut {NoStop}%
\bibitem [{\citenamefont {Lemke}\ \emph {et~al.}(1980)\citenamefont {Lemke},
  \citenamefont {Trinkl},\ and\ \citenamefont {Hasselmann}}]{LTH1980}%
  \BibitemOpen
  \bibfield  {author} {\bibinfo {author} {\bibfnamefont {P.}~\bibnamefont
  {Lemke}}, \bibinfo {author} {\bibfnamefont {E.}~\bibnamefont {Trinkl}}, \
  and\ \bibinfo {author} {\bibfnamefont {K.}~\bibnamefont {Hasselmann}},\
  }\href@noop {} {\bibfield  {journal} {\bibinfo  {journal} {Journal of
  Physical Oceanography}\ }\textbf {\bibinfo {volume} {10}},\ \bibinfo {pages}
  {2100} (\bibinfo {year} {1980})}\BibitemShut {NoStop}%
\bibitem [{\citenamefont {Nicolis}(1981)}]{N1981}%
  \BibitemOpen
  \bibfield  {author} {\bibinfo {author} {\bibfnamefont {C.}~\bibnamefont
  {Nicolis}},\ }in\ \href@noop {} {\emph {\bibinfo {booktitle} {Physics of
  Solar Variations}}}\ (\bibinfo  {publisher} {Springer},\ \bibinfo {year}
  {1981})\ pp.\ \bibinfo {pages} {473--478}\BibitemShut {NoStop}%
\bibitem [{\citenamefont {Nicolis}(1982)}]{N1982}%
  \BibitemOpen
  \bibfield  {author} {\bibinfo {author} {\bibfnamefont {C.}~\bibnamefont
  {Nicolis}},\ }\href@noop {} {\bibfield  {journal} {\bibinfo  {journal}
  {Tellus}\ }\textbf {\bibinfo {volume} {34}},\ \bibinfo {pages} {1} (\bibinfo
  {year} {1982})}\BibitemShut {NoStop}%
\bibitem [{\citenamefont {Penland}(1989)}]{P1989}%
  \BibitemOpen
  \bibfield  {author} {\bibinfo {author} {\bibfnamefont {C.}~\bibnamefont
  {Penland}},\ }\href@noop {} {\bibfield  {journal} {\bibinfo  {journal}
  {Monthly Weather Review}\ }\textbf {\bibinfo {volume} {117}},\ \bibinfo
  {pages} {2165} (\bibinfo {year} {1989})}\BibitemShut {NoStop}%
\bibitem [{\citenamefont {Arnold}\ \emph {et~al.}(2003)\citenamefont {Arnold},
  \citenamefont {Imkeller},\ and\ \citenamefont {Wu}}]{AIW2003}%
  \BibitemOpen
  \bibfield  {author} {\bibinfo {author} {\bibfnamefont {L.}~\bibnamefont
  {Arnold}}, \bibinfo {author} {\bibfnamefont {P.}~\bibnamefont {Imkeller}}, \
  and\ \bibinfo {author} {\bibfnamefont {Y.}~\bibnamefont {Wu}},\ }\href@noop
  {} {\bibfield  {journal} {\bibinfo  {journal} {Dynamical Systems}\ }\textbf
  {\bibinfo {volume} {18}},\ \bibinfo {pages} {295} (\bibinfo {year}
  {2003})}\BibitemShut {NoStop}%
\bibitem [{\citenamefont {Penland}\ and\ \citenamefont
  {Matrosova}(1994)}]{PM1994}%
  \BibitemOpen
  \bibfield  {author} {\bibinfo {author} {\bibfnamefont {C.}~\bibnamefont
  {Penland}}\ and\ \bibinfo {author} {\bibfnamefont {L.}~\bibnamefont
  {Matrosova}},\ }\href@noop {} {\bibfield  {journal} {\bibinfo  {journal}
  {Journal of climate}\ }\textbf {\bibinfo {volume} {7}},\ \bibinfo {pages}
  {1352} (\bibinfo {year} {1994})}\BibitemShut {NoStop}%
\bibitem [{\citenamefont {Penland}(1996)}]{P1996}%
  \BibitemOpen
  \bibfield  {author} {\bibinfo {author} {\bibfnamefont {C.}~\bibnamefont
  {Penland}},\ }\href@noop {} {\bibfield  {journal} {\bibinfo  {journal}
  {Physica D: Nonlinear Phenomena}\ }\textbf {\bibinfo {volume} {98}},\
  \bibinfo {pages} {534} (\bibinfo {year} {1996})}\BibitemShut {NoStop}%
\bibitem [{\citenamefont {Newman}\ \emph {et~al.}(1997)\citenamefont {Newman},
  \citenamefont {Sardeshmukh},\ and\ \citenamefont {Penland}}]{NSP1997}%
  \BibitemOpen
  \bibfield  {author} {\bibinfo {author} {\bibfnamefont {M.}~\bibnamefont
  {Newman}}, \bibinfo {author} {\bibfnamefont {P.~D.}\ \bibnamefont
  {Sardeshmukh}}, \ and\ \bibinfo {author} {\bibfnamefont {C.}~\bibnamefont
  {Penland}},\ }\href@noop {} {\bibfield  {journal} {\bibinfo  {journal}
  {Journal of the atmospheric sciences}\ }\textbf {\bibinfo {volume} {54}},\
  \bibinfo {pages} {435} (\bibinfo {year} {1997})}\BibitemShut {NoStop}%
\bibitem [{\citenamefont {Frederiksen}\ and\ \citenamefont
  {Davies}(1997)}]{FD1997}%
  \BibitemOpen
  \bibfield  {author} {\bibinfo {author} {\bibfnamefont {J.~S.}\ \bibnamefont
  {Frederiksen}}\ and\ \bibinfo {author} {\bibfnamefont {A.~G.}\ \bibnamefont
  {Davies}},\ }\href@noop {} {\bibfield  {journal} {\bibinfo  {journal}
  {Journal of the atmospheric sciences}\ }\textbf {\bibinfo {volume} {54}},\
  \bibinfo {pages} {2475} (\bibinfo {year} {1997})}\BibitemShut {NoStop}%
\bibitem [{\citenamefont {Frederiksen}(1999)}]{F1999}%
  \BibitemOpen
  \bibfield  {author} {\bibinfo {author} {\bibfnamefont {J.~S.}\ \bibnamefont
  {Frederiksen}},\ }\href@noop {} {\bibfield  {journal} {\bibinfo  {journal}
  {Journal of the atmospheric sciences}\ }\textbf {\bibinfo {volume} {56}},\
  \bibinfo {pages} {1481} (\bibinfo {year} {1999})}\BibitemShut {NoStop}%
\bibitem [{\citenamefont {Nicolis}(2003)}]{N2003}%
  \BibitemOpen
  \bibfield  {author} {\bibinfo {author} {\bibfnamefont {C.}~\bibnamefont
  {Nicolis}},\ }\href@noop {} {\bibfield  {journal} {\bibinfo  {journal}
  {Journal of the Atmospheric Sciences}\ }\textbf {\bibinfo {volume} {60}},\
  \bibinfo {pages} {2208} (\bibinfo {year} {2003})}\BibitemShut {NoStop}%
\bibitem [{\citenamefont {Nicolis}(2004)}]{N2004}%
  \BibitemOpen
  \bibfield  {author} {\bibinfo {author} {\bibfnamefont {C.}~\bibnamefont
  {Nicolis}},\ }\href@noop {} {\bibfield  {journal} {\bibinfo  {journal}
  {Journal of the Atmospheric Sciences}\ }\textbf {\bibinfo {volume} {61}},\
  \bibinfo {pages} {1740} (\bibinfo {year} {2004})}\BibitemShut {NoStop}%
\bibitem [{\citenamefont {Buizza}\ \emph {et~al.}(1999)\citenamefont {Buizza},
  \citenamefont {Milleer},\ and\ \citenamefont {Palmer}}]{BMP1999}%
  \BibitemOpen
  \bibfield  {author} {\bibinfo {author} {\bibfnamefont {R.}~\bibnamefont
  {Buizza}}, \bibinfo {author} {\bibfnamefont {M.}~\bibnamefont {Milleer}}, \
  and\ \bibinfo {author} {\bibfnamefont {T.}~\bibnamefont {Palmer}},\
  }\href@noop {} {\bibfield  {journal} {\bibinfo  {journal} {Quarterly Journal
  of the Royal Meteorological Society}\ }\textbf {\bibinfo {volume} {125}},\
  \bibinfo {pages} {2887} (\bibinfo {year} {1999})}\BibitemShut {NoStop}%
\bibitem [{\citenamefont {Shutts}(2005)}]{S2005}%
  \BibitemOpen
  \bibfield  {author} {\bibinfo {author} {\bibfnamefont {G.}~\bibnamefont
  {Shutts}},\ }\href@noop {} {\bibfield  {journal} {\bibinfo  {journal}
  {Quarterly Journal of the Royal Meteorological Society}\ }\textbf {\bibinfo
  {volume} {131}},\ \bibinfo {pages} {3079} (\bibinfo {year}
  {2005})}\BibitemShut {NoStop}%
\bibitem [{\citenamefont {Nicolis}(2005)}]{N2005}%
  \BibitemOpen
  \bibfield  {author} {\bibinfo {author} {\bibfnamefont {C.}~\bibnamefont
  {Nicolis}},\ }\href@noop {} {\bibfield  {journal} {\bibinfo  {journal}
  {Quarterly Journal of the Royal Meteorological Society}\ }\textbf {\bibinfo
  {volume} {131}},\ \bibinfo {pages} {2151} (\bibinfo {year}
  {2005})}\BibitemShut {NoStop}%
\bibitem [{\citenamefont {Doblas-Reyes}\ \emph {et~al.}(2009)\citenamefont
  {Doblas-Reyes}, \citenamefont {Weisheimer}, \citenamefont {D{\'e}qu{\'e}},
  \citenamefont {Keenlyside}, \citenamefont {McVean}, \citenamefont {Murphy},
  \citenamefont {Rogel}, \citenamefont {Smith},\ and\ \citenamefont
  {Palmer}}]{Da2009}%
  \BibitemOpen
  \bibfield  {author} {\bibinfo {author} {\bibfnamefont {F.}~\bibnamefont
  {Doblas-Reyes}}, \bibinfo {author} {\bibfnamefont {A.}~\bibnamefont
  {Weisheimer}}, \bibinfo {author} {\bibfnamefont {M.}~\bibnamefont
  {D{\'e}qu{\'e}}}, \bibinfo {author} {\bibfnamefont {N.}~\bibnamefont
  {Keenlyside}}, \bibinfo {author} {\bibfnamefont {M.}~\bibnamefont {McVean}},
  \bibinfo {author} {\bibfnamefont {J.}~\bibnamefont {Murphy}}, \bibinfo
  {author} {\bibfnamefont {P.}~\bibnamefont {Rogel}}, \bibinfo {author}
  {\bibfnamefont {D.}~\bibnamefont {Smith}}, \ and\ \bibinfo {author}
  {\bibfnamefont {T.}~\bibnamefont {Palmer}},\ }\href@noop {} {\bibfield
  {journal} {\bibinfo  {journal} {Quarterly Journal of the Royal Meteorological
  Society}\ }\textbf {\bibinfo {volume} {135}},\ \bibinfo {pages} {1538}
  (\bibinfo {year} {2009})}\BibitemShut {NoStop}%
\bibitem [{\citenamefont {Sura}\ \emph {et~al.}(2005)\citenamefont {Sura},
  \citenamefont {Newman}, \citenamefont {Penland},\ and\ \citenamefont
  {Sardeshmukh}}]{SNPS2005}%
  \BibitemOpen
  \bibfield  {author} {\bibinfo {author} {\bibfnamefont {P.}~\bibnamefont
  {Sura}}, \bibinfo {author} {\bibfnamefont {M.}~\bibnamefont {Newman}},
  \bibinfo {author} {\bibfnamefont {C.}~\bibnamefont {Penland}}, \ and\
  \bibinfo {author} {\bibfnamefont {P.}~\bibnamefont {Sardeshmukh}},\
  }\href@noop {} {\bibfield  {journal} {\bibinfo  {journal} {Journal of the
  atmospheric sciences}\ }\textbf {\bibinfo {volume} {62}},\ \bibinfo {pages}
  {1391} (\bibinfo {year} {2005})}\BibitemShut {NoStop}%
\bibitem [{\citenamefont {Sardeshmukh}\ and\ \citenamefont
  {Penland}(2015)}]{SP2015}%
  \BibitemOpen
  \bibfield  {author} {\bibinfo {author} {\bibfnamefont {P.~D.}\ \bibnamefont
  {Sardeshmukh}}\ and\ \bibinfo {author} {\bibfnamefont {C.}~\bibnamefont
  {Penland}},\ }\href@noop {} {\bibfield  {journal} {\bibinfo  {journal}
  {Chaos: An Interdisciplinary Journal of Nonlinear Science}\ }\textbf
  {\bibinfo {volume} {25}},\ \bibinfo {pages} {036410} (\bibinfo {year}
  {2015})}\BibitemShut {NoStop}%
\bibitem [{\citenamefont {Sura}(2013)}]{S2013}%
  \BibitemOpen
  \bibfield  {author} {\bibinfo {author} {\bibfnamefont {P.}~\bibnamefont
  {Sura}},\ }in\ \href@noop {} {\emph {\bibinfo {booktitle} {Extremes in a
  Changing Climate}}}\ (\bibinfo  {publisher} {Springer},\ \bibinfo {year}
  {2013})\ pp.\ \bibinfo {pages} {181--222}\BibitemShut {NoStop}%
\bibitem [{\citenamefont {Kifer}(2001)}]{K2001}%
  \BibitemOpen
  \bibfield  {author} {\bibinfo {author} {\bibfnamefont {Y.}~\bibnamefont
  {Kifer}},\ }in\ \href@noop {} {\emph {\bibinfo {booktitle} {Stochastic
  Climate Models}}}\ (\bibinfo  {publisher} {Springer},\ \bibinfo {year}
  {2001})\ pp.\ \bibinfo {pages} {171--188}\BibitemShut {NoStop}%
\bibitem [{\citenamefont {Arnold}(2001)}]{A2001}%
  \BibitemOpen
  \bibfield  {author} {\bibinfo {author} {\bibfnamefont {L.}~\bibnamefont
  {Arnold}},\ }in\ \href@noop {} {\emph {\bibinfo {booktitle} {Stochastic
  climate models}}}\ (\bibinfo  {publisher} {Springer},\ \bibinfo {year}
  {2001})\ pp.\ \bibinfo {pages} {141--157}\BibitemShut {NoStop}%
\bibitem [{\citenamefont {Kifer}(2003)}]{K2003}%
  \BibitemOpen
  \bibfield  {author} {\bibinfo {author} {\bibfnamefont {Y.}~\bibnamefont
  {Kifer}},\ }\href@noop {} {\bibfield  {journal} {\bibinfo  {journal}
  {Stochastics and Dynamics}\ }\textbf {\bibinfo {volume} {3}},\ \bibinfo
  {pages} {213} (\bibinfo {year} {2003})}\BibitemShut {NoStop}%
\bibitem [{\citenamefont {Majda}\ \emph {et~al.}(2001)\citenamefont {Majda},
  \citenamefont {Timofeyev},\ and\ \citenamefont {Vanden~Eijnden}}]{MTV2001}%
  \BibitemOpen
  \bibfield  {author} {\bibinfo {author} {\bibfnamefont {A.~J.}\ \bibnamefont
  {Majda}}, \bibinfo {author} {\bibfnamefont {I.}~\bibnamefont {Timofeyev}}, \
  and\ \bibinfo {author} {\bibfnamefont {E.}~\bibnamefont {Vanden~Eijnden}},\
  }\href@noop {} {\bibfield  {journal} {\bibinfo  {journal} {Communications on
  Pure and Applied Mathematics}\ }\textbf {\bibinfo {volume} {54}},\ \bibinfo
  {pages} {891} (\bibinfo {year} {2001})}\BibitemShut {NoStop}%
\bibitem [{\citenamefont {Franzke}\ \emph {et~al.}(2005)\citenamefont
  {Franzke}, \citenamefont {Majda},\ and\ \citenamefont
  {Vanden-Eijnden}}]{FMV2005}%
  \BibitemOpen
  \bibfield  {author} {\bibinfo {author} {\bibfnamefont {C.}~\bibnamefont
  {Franzke}}, \bibinfo {author} {\bibfnamefont {A.~J.}\ \bibnamefont {Majda}},
  \ and\ \bibinfo {author} {\bibfnamefont {E.}~\bibnamefont {Vanden-Eijnden}},\
  }\href@noop {} {\bibfield  {journal} {\bibinfo  {journal} {Journal of the
  atmospheric sciences}\ }\textbf {\bibinfo {volume} {62}},\ \bibinfo {pages}
  {1722} (\bibinfo {year} {2005})}\BibitemShut {NoStop}%
\bibitem [{\citenamefont {Culina}\ \emph {et~al.}(2011)\citenamefont {Culina},
  \citenamefont {Kravtsov},\ and\ \citenamefont {Monahan}}]{CKM2011}%
  \BibitemOpen
  \bibfield  {author} {\bibinfo {author} {\bibfnamefont {J.}~\bibnamefont
  {Culina}}, \bibinfo {author} {\bibfnamefont {S.}~\bibnamefont {Kravtsov}}, \
  and\ \bibinfo {author} {\bibfnamefont {A.~H.}\ \bibnamefont {Monahan}},\
  }\href@noop {} {\bibfield  {journal} {\bibinfo  {journal} {Journal of the
  Atmospheric Sciences}\ }\textbf {\bibinfo {volume} {68}},\ \bibinfo {pages}
  {284} (\bibinfo {year} {2011})}\BibitemShut {NoStop}%
\bibitem [{\citenamefont {Vannitsem}(2014)}]{V2014}%
  \BibitemOpen
  \bibfield  {author} {\bibinfo {author} {\bibfnamefont {S.}~\bibnamefont
  {Vannitsem}},\ }\href@noop {} {\bibfield  {journal} {\bibinfo  {journal}
  {Philosophical Transactions of the Royal Society of London A: Mathematical,
  Physical and Engineering Sciences}\ }\textbf {\bibinfo {volume} {372}},\
  \bibinfo {pages} {20130282} (\bibinfo {year} {2014})}\BibitemShut {NoStop}%
\bibitem [{\citenamefont {Freidlin}\ and\ \citenamefont
  {Wentzell}(1984)}]{FW1984}%
  \BibitemOpen
  \bibfield  {author} {\bibinfo {author} {\bibfnamefont {M.~I.}\ \bibnamefont
  {Freidlin}}\ and\ \bibinfo {author} {\bibfnamefont {A.~D.}\ \bibnamefont
  {Wentzell}},\ }in\ \href@noop {} {\emph {\bibinfo {booktitle} {Random
  Perturbations of Dynamical Systems}}}\ (\bibinfo  {publisher} {Springer},\
  \bibinfo {year} {1984})\ pp.\ \bibinfo {pages} {15--43}\BibitemShut {NoStop}%
\bibitem [{\citenamefont {Bouchet}\ \emph {et~al.}(2016)\citenamefont
  {Bouchet}, \citenamefont {Grafke}, \citenamefont {Tangarife},\ and\
  \citenamefont {Vanden-Eijnden}}]{BGTV2016}%
  \BibitemOpen
  \bibfield  {author} {\bibinfo {author} {\bibfnamefont {F.}~\bibnamefont
  {Bouchet}}, \bibinfo {author} {\bibfnamefont {T.}~\bibnamefont {Grafke}},
  \bibinfo {author} {\bibfnamefont {T.}~\bibnamefont {Tangarife}}, \ and\
  \bibinfo {author} {\bibfnamefont {E.}~\bibnamefont {Vanden-Eijnden}},\
  }\href@noop {} {\bibfield  {journal} {\bibinfo  {journal} {Journal of
  Statistical Physics}\ }\textbf {\bibinfo {volume} {162}},\ \bibinfo {pages}
  {793} (\bibinfo {year} {2016})}\BibitemShut {NoStop}%
\bibitem [{\citenamefont {Crommelin}\ and\ \citenamefont
  {Vanden-Eijnden}(2008)}]{CV2008}%
  \BibitemOpen
  \bibfield  {author} {\bibinfo {author} {\bibfnamefont {D.}~\bibnamefont
  {Crommelin}}\ and\ \bibinfo {author} {\bibfnamefont {E.}~\bibnamefont
  {Vanden-Eijnden}},\ }\href@noop {} {\bibfield  {journal} {\bibinfo  {journal}
  {Journal of the Atmospheric Sciences}\ }\textbf {\bibinfo {volume} {65}},\
  \bibinfo {pages} {2661} (\bibinfo {year} {2008})}\BibitemShut {NoStop}%
\bibitem [{\citenamefont {Abramov}(2015)}]{A2015}%
  \BibitemOpen
  \bibfield  {author} {\bibinfo {author} {\bibfnamefont {R.}~\bibnamefont
  {Abramov}},\ }\href@noop {} {\bibfield  {journal} {\bibinfo  {journal}
  {Fluids}\ }\textbf {\bibinfo {volume} {1}},\ \bibinfo {pages} {2} (\bibinfo
  {year} {2015})}\BibitemShut {NoStop}%
\bibitem [{\citenamefont {Wouters}\ and\ \citenamefont
  {Lucarini}(2012)}]{WL2012}%
  \BibitemOpen
  \bibfield  {author} {\bibinfo {author} {\bibfnamefont {J.}~\bibnamefont
  {Wouters}}\ and\ \bibinfo {author} {\bibfnamefont {V.}~\bibnamefont
  {Lucarini}},\ }\href@noop {} {\bibfield  {journal} {\bibinfo  {journal}
  {Journal of Statistical Mechanics: Theory and Experiment}\ }\textbf {\bibinfo
  {volume} {2012}},\ \bibinfo {pages} {P03003} (\bibinfo {year}
  {2012})}\BibitemShut {NoStop}%
\bibitem [{\citenamefont {Demaeyer}\ and\ \citenamefont
  {Vannitsem}(2016)}]{DV2016}%
  \BibitemOpen
  \bibfield  {author} {\bibinfo {author} {\bibfnamefont {J.}~\bibnamefont
  {Demaeyer}}\ and\ \bibinfo {author} {\bibfnamefont {S.}~\bibnamefont
  {Vannitsem}},\ }\href {\doibase 10.1002/qj.2973} {\bibfield  {journal}
  {\bibinfo  {journal} {Quarterly Journal of the Royal Meteorological Society,
  in press}\ } (\bibinfo {year} {2016}),\ 10.1002/qj.2973}\BibitemShut
  {NoStop}%
\bibitem [{\citenamefont {Wouters}\ \emph {et~al.}(2016)\citenamefont
  {Wouters}, \citenamefont {Dolaptchiev}, \citenamefont {Lucarini},\ and\
  \citenamefont {Achatz}}]{WDLA2016}%
  \BibitemOpen
  \bibfield  {author} {\bibinfo {author} {\bibfnamefont {J.}~\bibnamefont
  {Wouters}}, \bibinfo {author} {\bibfnamefont {S.~I.}\ \bibnamefont
  {Dolaptchiev}}, \bibinfo {author} {\bibfnamefont {V.}~\bibnamefont
  {Lucarini}}, \ and\ \bibinfo {author} {\bibfnamefont {U.}~\bibnamefont
  {Achatz}},\ }\href@noop {} {\bibfield  {journal} {\bibinfo  {journal}
  {Nonlinear Processes in Geophysics}\ }\textbf {\bibinfo {volume} {23}},\
  \bibinfo {pages} {435} (\bibinfo {year} {2016})}\BibitemShut {NoStop}%
\bibitem [{\citenamefont {Vissio}\ and\ \citenamefont
  {Lucarini}(2016)}]{VL2016}%
  \BibitemOpen
  \bibfield  {author} {\bibinfo {author} {\bibfnamefont {G.}~\bibnamefont
  {Vissio}}\ and\ \bibinfo {author} {\bibfnamefont {V.}~\bibnamefont
  {Lucarini}},\ }\href@noop {} {\bibfield  {journal} {\bibinfo  {journal}
  {arXiv preprint arXiv:1612.07223}\ } (\bibinfo {year} {2016})}\BibitemShut
  {NoStop}%
\bibitem [{\citenamefont {Sanders}\ and\ \citenamefont
  {Verhulst}(1985)}]{SV1985}%
  \BibitemOpen
  \bibfield  {author} {\bibinfo {author} {\bibfnamefont {J.}~\bibnamefont
  {Sanders}}\ and\ \bibinfo {author} {\bibfnamefont {F.}~\bibnamefont
  {Verhulst}},\ }\href@noop {} {\bibfield  {journal} {\bibinfo  {journal}
  {Applied Mathematical Sciences}\ }\textbf {\bibinfo {volume} {59}} (\bibinfo
  {year} {1985})}\BibitemShut {NoStop}%
\bibitem [{\citenamefont {Bogoliubov}\ and\ \citenamefont
  {Mitropolski}(1961)}]{BM1961}%
  \BibitemOpen
  \bibfield  {author} {\bibinfo {author} {\bibfnamefont {N.~N.}\ \bibnamefont
  {Bogoliubov}}\ and\ \bibinfo {author} {\bibfnamefont {Y.~A.}\ \bibnamefont
  {Mitropolski}},\ }\href@noop {} {\bibfield  {journal} {\bibinfo  {journal}
  {Asymptotic Methods in the Theory of Non-Linear Oscillations, by NN
  Bogoliubov and YA Mitropolski. New York: Gordon and Breach, 1961.}\ }\textbf
  {\bibinfo {volume} {1}} (\bibinfo {year} {1961})}\BibitemShut {NoStop}%
\bibitem [{\citenamefont {Arnold}\ \emph {et~al.}(2013)\citenamefont {Arnold},
  \citenamefont {Moroz},\ and\ \citenamefont {Palmer}}]{AMP2013}%
  \BibitemOpen
  \bibfield  {author} {\bibinfo {author} {\bibfnamefont {H.}~\bibnamefont
  {Arnold}}, \bibinfo {author} {\bibfnamefont {I.}~\bibnamefont {Moroz}}, \
  and\ \bibinfo {author} {\bibfnamefont {T.}~\bibnamefont {Palmer}},\
  }\href@noop {} {\bibfield  {journal} {\bibinfo  {journal} {Philosophical
  Transactions of the Royal Society of London A: Mathematical, Physical and
  Engineering Sciences}\ }\textbf {\bibinfo {volume} {371}},\ \bibinfo {pages}
  {20110479} (\bibinfo {year} {2013})}\BibitemShut {NoStop}%
\bibitem [{\citenamefont {Ruelle}(1997)}]{R1997}%
  \BibitemOpen
  \bibfield  {author} {\bibinfo {author} {\bibfnamefont {D.}~\bibnamefont
  {Ruelle}},\ }\href@noop {} {\bibfield  {journal} {\bibinfo  {journal}
  {Communications in Mathematical Physics}\ }\textbf {\bibinfo {volume}
  {187}},\ \bibinfo {pages} {227} (\bibinfo {year} {1997})}\BibitemShut
  {NoStop}%
\bibitem [{\citenamefont {Ruelle}(2009)}]{R2009}%
  \BibitemOpen
  \bibfield  {author} {\bibinfo {author} {\bibfnamefont {D.}~\bibnamefont
  {Ruelle}},\ }\href@noop {} {\bibfield  {journal} {\bibinfo  {journal}
  {Nonlinearity}\ }\textbf {\bibinfo {volume} {22}},\ \bibinfo {pages} {855}
  (\bibinfo {year} {2009})}\BibitemShut {NoStop}%
\bibitem [{\citenamefont {Wouters}\ and\ \citenamefont
  {Lucarini}(2013)}]{WL2013}%
  \BibitemOpen
  \bibfield  {author} {\bibinfo {author} {\bibfnamefont {J.}~\bibnamefont
  {Wouters}}\ and\ \bibinfo {author} {\bibfnamefont {V.}~\bibnamefont
  {Lucarini}},\ }\href@noop {} {\bibfield  {journal} {\bibinfo  {journal}
  {Journal of Statistical Physics}\ }\textbf {\bibinfo {volume} {151}},\
  \bibinfo {pages} {850} (\bibinfo {year} {2013})}\BibitemShut {NoStop}%
\bibitem [{\citenamefont {Young}(2002)}]{Y2002}%
  \BibitemOpen
  \bibfield  {author} {\bibinfo {author} {\bibfnamefont {L.-S.}\ \bibnamefont
  {Young}},\ }\href@noop {} {\bibfield  {journal} {\bibinfo  {journal} {Journal
  of Statistical Physics}\ }\textbf {\bibinfo {volume} {108}},\ \bibinfo
  {pages} {733} (\bibinfo {year} {2002})}\BibitemShut {NoStop}%
\bibitem [{\citenamefont {Chekroun}\ \emph {et~al.}(2015)\citenamefont
  {Chekroun}, \citenamefont {Liu},\ and\ \citenamefont {Wang}}]{CLW2015}%
  \BibitemOpen
  \bibfield  {author} {\bibinfo {author} {\bibfnamefont {M.~D.}\ \bibnamefont
  {Chekroun}}, \bibinfo {author} {\bibfnamefont {H.}~\bibnamefont {Liu}}, \
  and\ \bibinfo {author} {\bibfnamefont {S.}~\bibnamefont {Wang}},\ }\href@noop
  {} {\emph {\bibinfo {title} {{Approximation} of {Stochastic} {Invariant}
  {Manifolds:} {Stochastic} {Manifolds} for {Nonlinear} {SPDEs I}}}}\ (\bibinfo
   {publisher} {Springer},\ \bibinfo {year} {2015})\BibitemShut {NoStop}%
\bibitem [{\citenamefont {Grad}(1969)}]{G1969}%
  \BibitemOpen
  \bibfield  {author} {\bibinfo {author} {\bibfnamefont {H.}~\bibnamefont
  {Grad}},\ }\href@noop {} {\bibfield  {journal} {\bibinfo  {journal}
  {Transport theory}\ }\textbf {\bibinfo {volume} {1}},\ \bibinfo {pages} {269}
  (\bibinfo {year} {1969})}\BibitemShut {NoStop}%
\bibitem [{\citenamefont {Ellis}\ and\ \citenamefont {Pinsky}(1975)}]{EP1975}%
  \BibitemOpen
  \bibfield  {author} {\bibinfo {author} {\bibfnamefont {R.~S.}\ \bibnamefont
  {Ellis}}\ and\ \bibinfo {author} {\bibfnamefont {M.~A.}\ \bibnamefont
  {Pinsky}},\ }\href@noop {} {\bibfield  {journal} {\bibinfo  {journal} {J.
  Math. Pures Appl}\ }\textbf {\bibinfo {volume} {54}},\ \bibinfo {pages} {125}
  (\bibinfo {year} {1975})}\BibitemShut {NoStop}%
\bibitem [{\citenamefont {Papanicolaou}(1976)}]{P1976}%
  \BibitemOpen
  \bibfield  {author} {\bibinfo {author} {\bibfnamefont {G.~C.}\ \bibnamefont
  {Papanicolaou}},\ }\href@noop {} {\bibfield  {journal} {\bibinfo  {journal}
  {Journal of Mathematics}\ }\textbf {\bibinfo {volume} {6}} (\bibinfo {year}
  {1976})}\BibitemShut {NoStop}%
\bibitem [{\citenamefont {Kurtz}(1973)}]{K1973}%
  \BibitemOpen
  \bibfield  {author} {\bibinfo {author} {\bibfnamefont {T.~G.}\ \bibnamefont
  {Kurtz}},\ }\href@noop {} {\bibfield  {journal} {\bibinfo  {journal} {Journal
  of Functional Analysis}\ }\textbf {\bibinfo {volume} {12}},\ \bibinfo {pages}
  {55} (\bibinfo {year} {1973})}\BibitemShut {NoStop}%
\bibitem [{\citenamefont {Abramov}(2013)}]{A2013}%
  \BibitemOpen
  \bibfield  {author} {\bibinfo {author} {\bibfnamefont {R.~V.}\ \bibnamefont
  {Abramov}},\ }\href@noop {} {\bibfield  {journal} {\bibinfo  {journal}
  {Multiscale Modeling \& Simulation}\ }\textbf {\bibinfo {volume} {11}},\
  \bibinfo {pages} {134} (\bibinfo {year} {2013})}\BibitemShut {NoStop}%
\bibitem [{\citenamefont {Waleffe}(1992)}]{W1992}%
  \BibitemOpen
  \bibfield  {author} {\bibinfo {author} {\bibfnamefont {F.}~\bibnamefont
  {Waleffe}},\ }\href@noop {} {\bibfield  {journal} {\bibinfo  {journal}
  {Physics of Fluids A: Fluid Dynamics (1989-1993)}\ }\textbf {\bibinfo
  {volume} {4}},\ \bibinfo {pages} {350} (\bibinfo {year} {1992})}\BibitemShut
  {NoStop}%
\bibitem [{\citenamefont {Ohkitani}\ and\ \citenamefont {Kida}(1992)}]{OK1992}%
  \BibitemOpen
  \bibfield  {author} {\bibinfo {author} {\bibfnamefont {K.}~\bibnamefont
  {Ohkitani}}\ and\ \bibinfo {author} {\bibfnamefont {S.}~\bibnamefont
  {Kida}},\ }\href@noop {} {\bibfield  {journal} {\bibinfo  {journal} {Physics
  of Fluids A: Fluid Dynamics (1989-1993)}\ }\textbf {\bibinfo {volume} {4}},\
  \bibinfo {pages} {794} (\bibinfo {year} {1992})}\BibitemShut {NoStop}%
\bibitem [{\citenamefont {Smith}\ and\ \citenamefont {Waleffe}(1999)}]{SW1999}%
  \BibitemOpen
  \bibfield  {author} {\bibinfo {author} {\bibfnamefont {L.~M.}\ \bibnamefont
  {Smith}}\ and\ \bibinfo {author} {\bibfnamefont {F.}~\bibnamefont
  {Waleffe}},\ }\href@noop {} {\bibfield  {journal} {\bibinfo  {journal}
  {Physics of fluids}\ }\textbf {\bibinfo {volume} {11}},\ \bibinfo {pages}
  {1608} (\bibinfo {year} {1999})}\BibitemShut {NoStop}%
\bibitem [{\citenamefont {Hansen}\ and\ \citenamefont
  {Penland}(2006)}]{HP2006}%
  \BibitemOpen
  \bibfield  {author} {\bibinfo {author} {\bibfnamefont {J.~A.}\ \bibnamefont
  {Hansen}}\ and\ \bibinfo {author} {\bibfnamefont {C.}~\bibnamefont
  {Penland}},\ }\href@noop {} {\bibfield  {journal} {\bibinfo  {journal}
  {Monthly weather review}\ }\textbf {\bibinfo {volume} {134}},\ \bibinfo
  {pages} {3006} (\bibinfo {year} {2006})}\BibitemShut {NoStop}%
\bibitem [{\citenamefont {R{\"u}emelin}(1982)}]{R1982}%
  \BibitemOpen
  \bibfield  {author} {\bibinfo {author} {\bibfnamefont {W.}~\bibnamefont
  {R{\"u}emelin}},\ }\href@noop {} {\bibfield  {journal} {\bibinfo  {journal}
  {SIAM Journal on Numerical Analysis}\ }\textbf {\bibinfo {volume} {19}},\
  \bibinfo {pages} {604} (\bibinfo {year} {1982})}\BibitemShut {NoStop}%
\bibitem [{\citenamefont {Abramov}(2012)}]{A2012}%
  \BibitemOpen
  \bibfield  {author} {\bibinfo {author} {\bibfnamefont {R.~V.}\ \bibnamefont
  {Abramov}},\ }\href@noop {} {\bibfield  {journal} {\bibinfo  {journal}
  {Multiscale Modeling \& Simulation}\ }\textbf {\bibinfo {volume} {10}},\
  \bibinfo {pages} {28} (\bibinfo {year} {2012})}\BibitemShut {NoStop}%
\bibitem [{\citenamefont {Gardiner}(2009)}]{G2009}%
  \BibitemOpen
  \bibfield  {author} {\bibinfo {author} {\bibfnamefont {C.~W.}\ \bibnamefont
  {Gardiner}},\ }\href@noop {} {\emph {\bibinfo {title} {Handbook of stochastic
  methods}}},\ \bibinfo {edition} {4th}\ ed.\ (\bibinfo  {publisher} {Springer
  Berlin},\ \bibinfo {year} {2009})\BibitemShut {NoStop}%
\bibitem [{\citenamefont {De~Cruz}\ \emph {et~al.}(2016)\citenamefont
  {De~Cruz}, \citenamefont {Demaeyer},\ and\ \citenamefont
  {Vannitsem}}]{DDV2016}%
  \BibitemOpen
  \bibfield  {author} {\bibinfo {author} {\bibfnamefont {L.}~\bibnamefont
  {De~Cruz}}, \bibinfo {author} {\bibfnamefont {J.}~\bibnamefont {Demaeyer}}, \
  and\ \bibinfo {author} {\bibfnamefont {S.}~\bibnamefont {Vannitsem}},\
  }\href@noop {} {\bibfield  {journal} {\bibinfo  {journal} {Geoscientific
  Model Development}\ }\textbf {\bibinfo {volume} {9}},\ \bibinfo {pages}
  {2793} (\bibinfo {year} {2016})}\BibitemShut {NoStop}%
\end{thebibliography}%

\end{document}